\documentclass[%
reprint,
superscriptaddress,
nofootinbib,
nobibnotes,
 amsmath,amssymb,
 aps,
prl,
]{revtex4-2}
\usepackage[dvipsnames]{xcolor}
\usepackage{graphicx}
\usepackage{dcolumn}
\usepackage{bm}
\usepackage[breaklinks=true,colorlinks,citecolor=blue,linkcolor=red,urlcolor=blue]{hyperref}

\usepackage{braket}
\usepackage[utf8]{inputenc} 
\usepackage{mathtools}
\usepackage[bottom]{footmisc}
\usepackage{hyperref}
\usepackage{amsfonts}
\usepackage{amsmath}
\usepackage{slashed}
\usepackage{amsthm}
\usepackage{mathbbol}
\usepackage{soul}

\usepackage{tikz}

\begin{document}


\title{Extending Topological Bound on Quantum Weight Beyond Symmetry-Protected Topological Phases}

\author{Yi-Chun Hung$\,^{\hyperlink{equal}{*},\hyperlink{email1}{\dagger}}$}
\affiliation{Department of Physics,\;Northeastern\;University,\;Boston,\;Massachusetts\,\;02115,\;USA}
\affiliation{Quantum Materials and Sensing Institute,\;Northeastern University,\;Burlington,\;Massachusetts\;01803,\;USA}

\author{Yugo Onishi$\,^{\hyperlink{equal}{*}}$}
\affiliation{Department of Physics, Massachusetts Institute of Technology, Cambridge, Massachusetts 02139, USA}

\author{Hsin Lin}
\affiliation{ Institute of Physics,\;Academia Sinica,\;Taipei,\;115201,\;Taiwan}

\author{Liang Fu}
\affiliation{Department of Physics, Massachusetts Institute of Technology, Cambridge, Massachusetts 02139, USA}

\author{Arun Bansil$\,^{\hyperlink{email2}{\ddagger}}$}
\affiliation{Department of Physics,\;Northeastern\;University,\;Boston,\;Massachusetts\,\;02115,\;USA}
\affiliation{Quantum Materials and Sensing Institute,\;Northeastern University,\;Burlington,\;Massachusetts\;01803,\;USA}


\begin{abstract}
The quantum metric encodes the geometric structure of Bloch wave functions and governs a wide range of physical responses. Its Brillouin-zone integral, the quantum weight, appears in the structure factor and provides lower bounds on observables such as the optical gap and dielectric constant. In symmetry-protected topological (SPT) phases, the nontrivial band topology imposes a lower bound on the quantum weight and constraints on the observables. Here, we generalize the topological bound on quantum geometry to encompass systems beyond the SPT phases. We show that topological invariants defined via the projected spectrum lower-bound the quantum weight with a symmetry-breaking correction to the quantum metric. Our proposed bound holds even when the underlying symmetries are broken, and it would be amenable to experimental verification via the optical conductivity sum rule under external fields. We illustrate our theory by adding a nonzero spin-orbit coupling term to a spin Chern insulator model, where we show that our proposed bound applies even though the conventional topological bound does not hold.
\end{abstract}

\maketitle
\renewcommand{\thefootnote}{\fnsymbol{footnote}}
\footnotetext[1]{\hypertarget{equal}{These authors contributed equally.}}
\footnotetext[2]{\hypertarget{email1}{Contact author: \href{mailto:hung.yi@northeastern.edu}{hung.yi@northeastern.edu}}}
\footnotetext[3]{\hypertarget{email2}{Contact author: \href{mailto:ar.bansil@northeastern.edu}{ar.bansil@northeastern.edu}}}
\paragraph*{\textbf{Introduction}---}Symmetry-protected topological (SPT) phases have profoundly shaped our understanding of quantum matter and helped establish a robust framework for understanding nontrivial band topologies \cite{RevModPhys.88.021004, PhysRevB.74.195312, PhysRevB.75.121306, PhysRevB.76.045302, PhysRevB.78.195424,Po2017, Bradlyn2017-pj, PhysRevX.7.041069}. Perturbations and interactions inevitably break symmetries, and raise the fundamental question as to how topological characteristics persist or evolve beyond the conventional SPT paradigm. Recent advances in spin-resolved topology provide a natural extension of SPT concepts using the projected spectrum \cite{PhysRevB.80.125327, Lin2024, shulman2010robust} to enable characterization of topological phases even when the symmetry is broken. Generalization of these ideas to encompass the projected spectrum of translationally invariant operators has revealed that band topology, as well as the associated bulk-boundary correspondence, can remain well-defined in the absence of explicit symmetry protection \cite{PhysRevLett.108.196806, YANG2018723, wang2023feature, PhysRevB.109.155143, PhysRevB.111.195102}. Despite this progress, the effects of broken symmetries on quantum geometry and the related physical responses in materials remain relatively unexplored. 

\par In this connection, quantum metric, $g^{\mu\nu}$, has emerged as a key property for quantifying the overlap of Bloch wave functions in the Brillouin zone (BZ) \cite{PhysRevB.56.12847}. Integral of $g^{\mu\nu}$ over the BZ, called the quantum weight~\cite{PhysRevResearch.7.023158}, quantifies the quantum geometry encoded in the quantum geometric tensor, $G^{\mu\nu}$, whose real and imaginary parts correspond to $g^{\mu\nu}$ and Berry curvature, $\Omega^{\mu\nu}$, respectively \cite{Provost1980} (See Eq.~\eqref{eq:G_def} below for their definitions). The quantum weight, $K^{\mu\nu}$, is defined as:
\begin{equation}\label{eq:K}
    K^{\mu\nu} = 2\pi \int d[\mathbf{k}] \, g^{\mu\nu},
\end{equation}
where $d[\mathbf{k}]=d^dk/(2\pi)^d$ is the integral measure in $d$-dimensional BZ. More generally, the quantum weight is defined for interacting systems and related to the structure factor~\cite{PhysRevResearch.7.023158}.

The quantum weight of gapped systems was shown to satisfy topological bounds for general interacting systems~\cite{PhysRevLett.133.206602, PhysRevX.14.011052}. For noninteracting systems, the bound can be understood through the relationships between $g^{\mu\nu}$, $\Omega^{\mu\nu}$, and the Wilson-loop winding, topological invariants such as the Chern number \cite{PhysRevB.90.165139, PhysRevB.95.024515, Peotta2015} and the $\mathbb{Z}_2$ index \cite{PhysRevLett.135.086401, yu2025wilsonloop} impose a lower bound on the trace of the quantum weight, $K=\text{tr}[K^{\mu\nu}]$, where tr[...] refers to trace over spatial indices. Such a topological bound also applies to systems carrying Chern numbers protected by additional symmetries, such as the spin-$U(1)$ symmetry, and leads to $K\geq\sum_{\alpha=\uparrow,\downarrow}|C_\alpha|$ \cite{PhysRevX.14.011052}. While this relation provides insight into symmetric systems, its applicability to symmetry-breaking phases has, to our knowledge, not been reported \cite{footnote}\nocite{SM, PhysRevR.7.L042011}. 

\par The quantum metric $g^{\mu\nu}$ and quantum weight $K$ capture the quantum geometric contributions to responses~\cite{Yu2025}, such as optical conductivity \cite{doi:10.1126/science.adf1506, doi:10.1126/sciadv.ado1761, Wang2023} and superfluid weight \cite{PhysRevB.95.024515}, especially in flat-band systems \cite{Peotta2015, PhysRevLett.124.167002, PhysRevLett.128.087002}. Moreover, through the optical sum rules, $K$ imposes bounds on physical observables, including upper bounds on the optical gap~\cite{PhysRevX.14.011052,Resta_2002} and static dielectric constant~\cite{PhysRevResearch.7.023158}, as well as a lower bound on the structure factor~\cite{PhysRevLett.133.206602,PhysRevB.112.035158}. The effect of symmetry breaking on the lower bound of $K$ is thus essential for refining these constraints in topological materials with broken symmetries. 

\par Here, we discuss how symmetry breaking alters quantum geometry via the projected spectrum and thereby affects the topological bound of the trace of the quantum weight $K$. Our main result is:
\begin{equation}\label{eq:result}
    K + K_c \geq \sum_\alpha|C_\alpha|, 
\end{equation}
where the right-hand side defines the topological bound on the Chern numbers, $C_\alpha$, of various symmetry sectors in the projected spectrum, and $K_c\geq0$ is the quantum geometric correction from the symmetry-breaking perturbations. Notably, since $K_c$ can be related to the optical conductivity sum rule under application of external fields, the bound in Eq.~\eqref{eq:result} could, in principle, be tested. 

For example, consider a spin Chern insulator with spin-$U(1)$ symmetry and spin Chern number $C_s=(C_{\uparrow}-C_{\downarrow})/2=\pm2$. In this case, the $\mathbb{Z}_2$ invariant is trivial, but the spin Chern number imposes the topological bound: $K\ge 2|C_s|=4$. When the system is perturbed so that the spin-$U(1)$ symmetry is broken, this topological bound no longer holds. However, our bound~\eqref{eq:result} remains valid. In this case, $K_c$ is generally finite, and the Chern number for each spin sector can be defined and remains equal to 2, yielding $K+K_c\ge 4$.

We demonstrate our theory by considering a spin Chern insulator (SCI) model and adding a spin-orbit coupling (SOC) to break the spin-$U(1)$ symmetry.  The conventional topological bound does not apply to the system with non-zero SOC, which is no longer in the SCI phase, but we show that the bound in Eq.~\eqref{eq:result} still holds.

\paragraph*{\textbf{Quantum geometry via the projected spectrum}---}In SCIs with broken spin-$U(1)$ symmetry, the spin Chern number can be generalized by projecting the occupied Bloch states onto the spin sectors through the spectrum of a projected spin operator $P\mathbf{\hat{S}}\cdot\mathbf{\hat{n}}P$ \cite{PhysRevB.80.125327}, where $P$ is the projection operator for the relevant states and $\mathbf{\hat{n}}$ is the directional vector of interest. One thus obtains a well-defined topological invariant beyond the conventional symmetry constraints, leading to the concept of spin-resolved topology \cite{PhysRevB.80.125327, Lin2024}. This approach can be extended to a translationally invariant operator $\hat{O}$ \cite{PhysRevB.80.125327, wang2023feature}, such as the mirror operator \cite{Liu2014, Hsieh2012} and the orbital angular momentum \cite{PhysRevB.111.195102, PhysRevLett.126.056601, Yao_2026}. This framework is based on the spectrum of the following projected operator:
\begin{equation}\label{eq:project_o}
    \mathcal{\hat{O}_{P}}=P\hat{O}P.
\end{equation}
We will hereon refer to the spectrum of $\mathcal{\hat{O}_{P}}$ as the \emph{projected spectrum}.

Since $\hat{O}$ respects translational symmetry, we can diagonalize $\mathcal{\hat{O}_{P}}$ and study its eigenvalue spectrum using Bloch's theorem \cite{wang2023feature}. 
To this end, we denote the eigenstates and the eigenvalues of $\mathcal{\hat{O}_{P}}$ as $\ket{\tilde{u}_n}$ and $\lambda_n$: 
\begin{equation}
    \mathcal{\hat{O}_{P}}\ket{\tilde{u}_n}=\lambda_{n}\ket{\tilde{u}_n}.
\end{equation}
Noting that $\hat{\mathcal{O}}_{P}$ acts nontrivially only on the occupied states, we expand the $\hat{\mathcal{O}}_{P}$ eigenstates $\ket{\tilde{u}_n}$ in terms of the occupied Bloch wave functions as
\begin{equation}\label{eq:U}
    \ket{\tilde{u}_n} = \sum_{m}U_{nm}\ket{u_m}.
\end{equation}
Here, $\ket{u_n}$ and $\ket{\tilde{u}_n}$ are the cell-periodic parts of the Bloch wave functions of the energy and $\hat{\mathcal{O}}_{P}$ eigenstates, respectively. Since $\ket{\tilde{u}_n}$ span the occupied subspace, the projector can be decomposed as 
\begin{equation}
P=\sum_{\alpha}P_{\alpha},
\end{equation}
where the projectors $P_\alpha$ are defined below.

\par
\begin{figure}[h]
  \centering
  \centering
    \includegraphics[width=\linewidth]{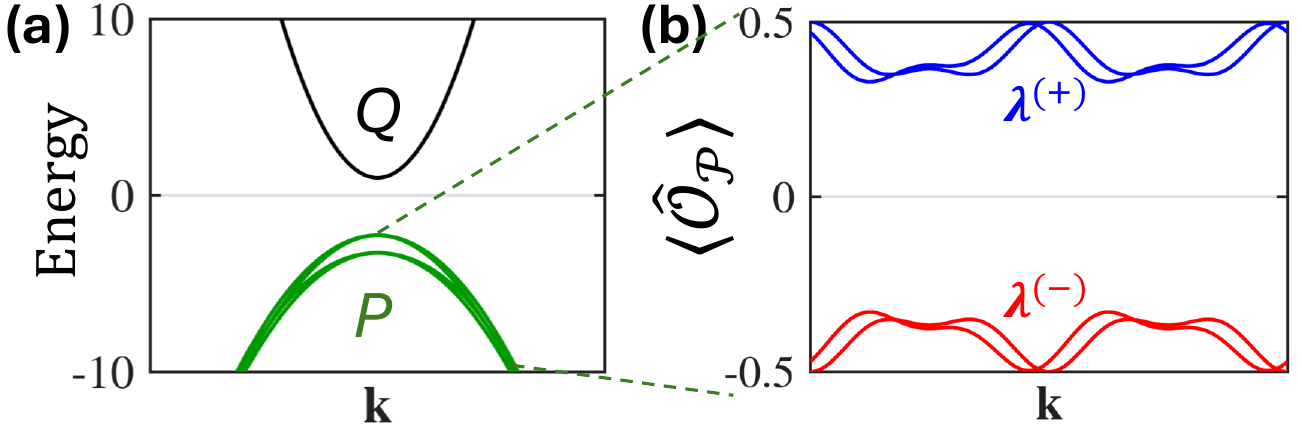}
  \caption{ Schematics of (a) band structure of an insulator and (b) various sectors categorized by $\lambda_n^{(\pm)}$ in the spectrum of $\mathcal{\hat{O}_{P}}$ (projected spectrum). Here, $Q=1-P$.
  }
  \label{fig:spectrum}
\end{figure}

 We group the $\ket{\tilde{u}_n}$ states into \emph{sectors} according to sets of isolated eigenvalues $\{\lambda_{n}^{(\alpha)}\}$, where $\alpha$ labels different sectors. The corresponding $\hat{\mathcal{O}}_{P}$ eigenstates are denoted as $\ket{\tilde{u}_n^{(\alpha)}}$, and the projector onto the $\alpha$th sector is $P_\alpha=\sum_{n\in\alpha}\ket{\tilde{u}_n^{(\alpha)}}\bra{\tilde{u}_n^{(\alpha)}}$. A schematic of the sectors in the projected spectrum is shown in Fig.~\ref{fig:spectrum}. Although we suppress $\vec{k}$ dependence, both the unitary transformation $U$ and the eigenvalues $\lambda_n$ generally depend on $\vec{k}$.

\par Having introduced the projected spectrum, we now turn to its quantum geometry. We first briefly revisit the quantum geometric tensor $G^{\mu\nu}$ \cite{Provost1980}:
\begin{equation}\label{eq:G_def}
G^{\mu\nu} = \text{Tr}[(\partial^\mu P)Q(\partial^\nu P)], 
\end{equation}
where $Q=\mathbf{1}-P$ and the trace (Tr$[...]$) sums over the band indices. It can be decomposed into $G^{\mu\nu}=g^{\mu\nu}-i\Omega^{\mu\nu}/2$, where $g^{\mu\nu}$ and $\Omega^{\mu\nu}$ are the quantum metric and Berry curvature of the relevant states, respectively \cite{Yu2025}. Furthermore, $G^{\mu\nu}$ can be interpreted as the Fubini-Study metric \cite{fubini1904sulle, Study1905} measuring the distance $dl$ between two relevant states that are infinitesimally separated in the BZ through \cite{PhysRevB.56.12847}
\begin{equation}\label{eq:dl}
    dl^2 = G^{\mu\nu}dk_{\mu}dk_{\nu} = \|QdP\|^2,
\end{equation}
where $\|O\| \equiv \sqrt{\text{Tr}[O^{\dagger}O]}$ is the Frobenius norm \cite{Horn_Johnson_1985} and $dP\equiv dk_{\mu} \partial^{\mu}P$. In the context of an operator inner product in the Hilbert space, $\langle O_1, O_2\rangle = \text{Tr}[O_1^{\dagger}O_2]$ \cite{conway1994course}, $QdP$ can be regarded as a ``vector" in the Hilbert space with length $dl$.

\par Applying the same construction to each sector in the projected spectrum, we define the sector-resolved quantum geometric tensor as
\begin{equation}\label{eq:Q_alpha}
G^{\mu\nu}_{\alpha} = \text{Tr}[(\partial^\mu P_{\alpha})Q_{\alpha}(\partial^\nu P_{\alpha})], 
\end{equation}
where $Q_{\alpha}=\mathbf{1}-P_{\alpha}$. The infinitesimal distance $dl_{\alpha}$ associated with $G^{\mu\nu}_{\alpha}$ is given by
\begin{align}\label{eq:alpha_dist}
    dl_{\alpha}^2 = G^{\mu\nu}_{\alpha}dk_{\mu}dk_{\nu} = \|Q_{\alpha}dP_{\alpha}\|^2,
\end{align}
where $dP_{\alpha}\equiv dk_{\mu} \partial^{\mu}P_{\alpha}$. $dl_{\alpha}$ defined in Eq.~\eqref{eq:alpha_dist} thus characterizes a sector-resolved distance between two states in the $\alpha$th sector that are infinitesimally separated in the BZ.

\paragraph*{\textbf{The topological bound beyond SPT phases}---} We now extend the topological bound on quantum weight beyond SPT phases by decomposing the quantum geometric tensor into sector-resolved contributions. Based on Eq.~\eqref{eq:alpha_dist}, $dl_{\alpha}$ can also be interpreted as the length of the ``vector" $Q_{\alpha}dP_{\alpha}$, which can be decomposed into two orthogonal components:
\begin{align}
    Q_{\alpha}dP_{\alpha} &= QdP_{\alpha} + (P-P_\alpha)dP_{\alpha}.
\end{align}
Since $\langle QdP_{\alpha},(P-P_\alpha)dP_{\alpha}\rangle=0$, it immediately implies a Pythagorean-type relation:
\begin{align}
    dl_\alpha^2 = dl_\alpha'^2 + dl_{c,\alpha}^2, \label{eq:pythagorean}
\end{align}
where $dl_\alpha'^2\equiv\|QdP_{\alpha}\|^2$ and $dl_{c}^2\equiv\|(P-P_\alpha)dP_{\alpha}\|^2$. When the $\hat{O}$ symmetry is preserved, $dl_{c}=0$ since the system is block-diagonalized so that $P_{\beta\neq\alpha}(\vec{k})P_\alpha(\vec{k}+\delta\vec{k})=0$ for any $\vec{k},\delta\vec{k}$, and thus $(P-P_{\alpha})dP_{\alpha}=0$. 
The nonzero $dl_{c,\alpha}$ captures the components along the ``inter-sector direction" introduced by symmetry breaking, in addition to the ``intra-sector direction" by $dl_\alpha'$.

Noting that $QdP=\sum_\alpha QdP_{\alpha}$ and $\langle QdP_{\alpha}, QdP_{\beta\neq\alpha}\rangle=0$, Pythagorean-type relation yields $\sum_{\alpha}dl_\alpha'^2=dl^2$. Therefore, summing over all the sectors in Eq.~\eqref{eq:pythagorean} yields
\begin{equation}\label{eq:pythagorean_2}
    \sum_{\alpha} dl_\alpha^2 = dl^2 + \sum_{\alpha}dl_{c,\alpha}^2.
\end{equation}
Consequently, in terms of the quantum geometric tensors, Eq.~\eqref{eq:pythagorean_2} yields
\begin{equation}\label{eq:G_decompose_2}
    \sum_{\alpha}G^{\mu\nu}_{\alpha} = G^{\mu\nu} + G^{\mu\nu}_{c},
\end{equation}
in which the correction term is
\begin{align}
G_{c}^{\mu\nu} & = \sum_{\alpha}G_{c,\alpha}^{\mu\nu},  \label{eq:G_c}
\\ G_{c,\alpha}^{\mu\nu} & = \text{Tr}[\partial^\mu P_\alpha(P-P_\alpha)\partial^\nu P_\alpha]. \label{eq:G_c_alpha}
\end{align}
Here, $G^{\mu\nu}_{\alpha}$, $G_{c,\alpha}^{\mu\nu}$, and, hence, $G_{c}^{\mu\nu}$ are Hermitian and positive semidefinite. Furthermore, $G_{c}^{\mu\nu}$ is real symmetric, see Supplemental Materials (SM) \cite{SM}. Through this decomposition, the total Berry curvature, $\Omega^{\mu\nu}$, thus equals the sum of the sector-resolved Berry curvatures, $\Omega^{\mu\nu}_\alpha$:
\begin{equation}
\Omega^{\mu\nu} = \sum_\alpha \Omega^{\mu\nu}_\alpha.
\end{equation}
In contrast, the total quantum metric, $g^{\mu\nu}$, with a correction term, equals the sum of the sector-resolved quantum metrics, $g^{\mu\nu}_\alpha$:
\begin{equation}\label{eq:metric_sum}
g^{\mu\nu} + G_{c}^{\mu\nu} = \sum_\alpha g^{\mu\nu}_\alpha, 
\end{equation}
where we have used the relations $g^{\mu\nu}_{(\alpha)}=\mathrm{Re}[G^{\mu\nu}_{(\alpha)}]$ and $\Omega^{\mu\nu}_{(\alpha)}=-2\mathrm{Im}[G^{\mu\nu}_{(\alpha)}]$ \cite{Yu2025}.

\par We now proceed to discuss our main result: the topological bound on quantum geometry persists beyond SPT phases. Following the steps of the earlier work on the Berry curvature bound of the quantum metric \cite{PhysRevB.90.165139, PhysRevB.95.024515, Peotta2015}, it is straightforward to show that the positive semidefiniteness of $G^{\mu\nu}_\alpha$ leads to
\begin{equation}\label{eq:05.5}
    \text{tr}[g_\alpha]\geq2\sqrt{\text{det}(g_\alpha)} \geq |\Omega_{\alpha}^{xy}|,
\end{equation}
where tr[...] sums over spatial indices. Similarly, from Eq.~\eqref{eq:metric_sum}, $K$ satisfies the inequality in Eq.~\eqref{eq:result}, in which $C_\alpha$ are sector-resolved Chern numbers and
\begin{equation}\label{eq:K_c}
K_c=2\pi\int d[\mathbf{k}] \text{tr}[G_c].
\end{equation}
Since $G_c$ is positive semidefinite, $K_c\geq0$. In view of Eq.~\eqref{eq:05.5}, the equality in Eq.~\eqref{eq:result} holds when $Q_\alpha(\partial^x+i\partial^y)\ket{\tilde{u}_{m}^{(\alpha)}}=0$ for all $m$ in every sector, as demonstrated with reasoning for positive semidefiniteness analogous to that for the total quantum metric \cite{PhysRevB.112.155158}. Note that $K_c$ is a generic correction to $K$ in the topological bound defined in Eq.~\eqref{eq:result}, arising from $\hat{O}$\text{-}symmetry breaking but admitting rare exceptions in which additional symmetries force $K_c=0$; such cases, however, are uncommon, so that $K_c$ remains a useful marker of $\hat{O}$\text{-}symmetry breaking, see SM for details \cite{SM}.

\begin{figure}[h]
  \centering
  \centering
    \includegraphics[width=\linewidth]{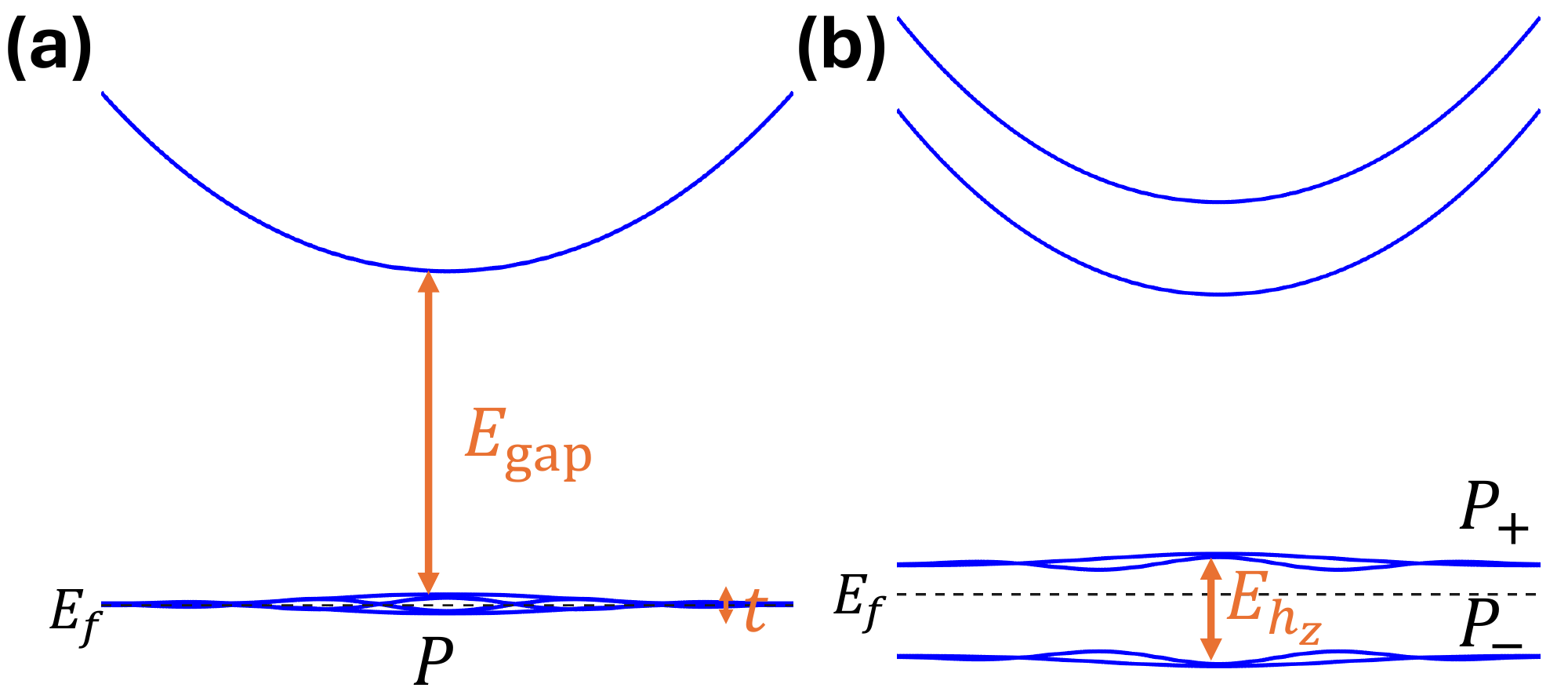}
  \caption{Schematics of the band structure of a SCI with narrow occupied-state bandwidth $t$ and a large band gap $E_{\text{gap}}$ (a) before and (b) after applying a Zeeman field with energy scale $E_{h_z}$. $E_f$ denotes the Fermi level, which is tuned to half-filling.}
  \label{fig:measure_Kc}
\end{figure}

\paragraph*{\textbf{Experimental implication of the topological bound }---} This section explains how $K_c$, and thus the topological bound in Eq.~\eqref{eq:result}, relates to the optical conductivity sum rule. As an illustrative example, we consider SCIs with a large band gap $E_{\text{gap}}$ and a narrow occupied-state bandwidth $t \ll E_{\text{gap}}$. The occupied states are assumed to lie near the Fermi level, making half filling readily accessible by, e.g., tuning the chemical potential (Fig.~\ref{fig:measure_Kc}(a)). We therefore consider the occupied states as the relevant states and $\hat{O} = \hat{S}^z$. To measure $K_c$, a Zeeman field $h_z$ is applied along the $z$-direction with energy scale $E_{h_z}$, which satisfies $t \ll E_{h_z} \ll E_{\text{gap}}$, and splits the occupied states across the Fermi level (Fig.~\ref{fig:measure_Kc}(b)). 

\par Since $E_{h_z} \ll E_{\text{gap}}$, the resulting low-energy Hamiltonian is:
\begin{equation}
H_{\text{low-E}} = P\left(H + h_z\hat{S}^z\right)P,
\end{equation}
where $P$ is the projection operator onto the occupied states and $H$ is the Hamiltonian of the SCI. Since $t \ll E_{h_z}$, we can approximate
\begin{equation}
H_{\text{low-E}} \approx P h_z\hat{S}^z P = h_z \mathcal{\hat{S}}^z_{\,\,\mathcal{P}}.
\end{equation}
Thus, $H_{\text{low-E}}$ and $\mathcal{\hat{S}}^z_{\,\,\mathcal{P}}$ share the same eigenstates, with the leading order corrections proportional to $\mathcal{O}(t/E_{h_z})$. The eigenstates with positive (negative) $\mathcal{\hat{S}}^z_{\,\,\mathcal{P}}$ eigenvalues therefore lie above (below) the Fermi level for $h_z > 0$ (Fig.~\ref{fig:measure_Kc}(b)). 

The optical transition between the states with positive and negative $\mathcal{\hat{S}}^z_{\,\,\mathcal{P}}$ eigenvalues is controlled by the sector-resolved quantum geometry. Choosing a cutoff frequency $\Omega$ satisfying $E_{h_z} < \hbar\Omega < E_{\text{gap}}$, the optical-conductivity sum rule yields \cite{PhysRevX.14.011052, PhysRevB.112.075116}
\begin{equation} \label{eq:sum_rule}
\int_0^\Omega d\omega \frac{\sigma^{(\text{abs})}_{\mu\nu}(\omega)}{\omega} = \frac{\pi e^2}{\hbar} \int d[\mathbf{k}] \text{Tr}\left[\partial^\mu P_- P_+ \partial^\nu P_-\right],
\end{equation}
where $\sigma^{(\text{abs})}_{\mu\nu}(\omega)$ denotes the absorptive part of conductivity and $P_+$ ($P_-$) is the projection operator onto states with positive (negative) $\mathcal{\hat{S}}^z_{\,\,\mathcal{P}}$ eigenvalues. From Eq.~\eqref{eq:G_c_alpha}, the integrand on the right-hand side of Eq.~\eqref{eq:sum_rule} is $G_{c,-}^{\mu\nu}$. Since there are only two sectors in the projected spectrum, $\text{tr}[G_{c,+}]=\text{tr}[G_{c,-}]$, see SM for details \cite{SM}. Consequently, by measuring $\text{tr}\left[\sigma^{(\text{abs})}(\omega)\right]$ using linearly polarized light, $K_c$ can be extracted via
\begin{equation}\label{eq:sum_rule_K_c}
K_c = \frac{4\hbar}{e^2} \int_0^\Omega d\omega \frac{\text{tr}\left[\sigma^{(\text{abs})}(\omega)\right]}{\omega}.
\end{equation}
$K$ can also be extracted via the optical conductivity sum rule before applying the Zeeman field \cite{PhysRevX.14.011052}: 
\begin{equation}\label{eq:sum_rule_K}
K = \frac{2\hbar}{e^2} \int_0^\infty d\omega \frac{\text{tr}\left[\sigma^{(\text{abs})}(\omega)\right]}{\omega}. 
\end{equation}
Combining Eqs.~\eqref{eq:sum_rule_K_c} and \eqref{eq:sum_rule_K}, the topological bound in Eq.~\eqref{eq:result} is thus experimentally verifiable via the optical conductivity sum rules by measuring $\text{tr}\left[\sigma^{(\text{abs})}(\omega)\right]$ using linearly polarized light. Additionally, in the present system, the topological bound in Eq.~\eqref{eq:result} can be traced back to the positivity of optical absorption power, see SM for details \cite{SM}.

\par Our conclusions can be generalized to systems satisfying $t \ll E_{\text{gap}}$ for certain $\hat{O}$s. Specifically, if $\hat{O}$ enters the Hamiltonian through a tunable physical field $F_{\hat{O}}$ and $\hat{\mathcal{O}}_P$ exhibits only two sectors, $K_c$ can be measured as follows. Tune $F_{\hat{O}}$ such that its associated energy scale $E_{F_{\hat{O}}}$ satisfies $t \ll E_{F_{\hat{O}}} \ll E_{\text{gap}}$, ensuring that eigenstates with opposite eigenvalues of $\hat{\mathcal{O}}_P$ lie on opposite sides of the Fermi level. Then, compute the optical conductivity sum rule in Eq.~\eqref{eq:sum_rule_K_c}, with some $\Omega$ satisfying $E_{F_{\hat{O}}} < \hbar\Omega < E_{\text{gap}}$. Together with the measurement of $K$ via Eq.~\eqref{eq:sum_rule_K} without applying the $F_{\hat{O}}$ field, the topological bound can be experimentally verified. A representative example of such a generalization is a bilayer of coupled Chern insulators, where $\hat{O}$ corresponds to the layer pseudospin and $F_{\hat{O}}$ is an out-of-plane electric field. The preceding analysis can be generalized to encompass $\hat{\mathcal{O}}_P$s with more sectors, see SM for details \cite{SM}.
  
\begin{figure}[h]
  \centering
  \centering
    \includegraphics[width=\linewidth]{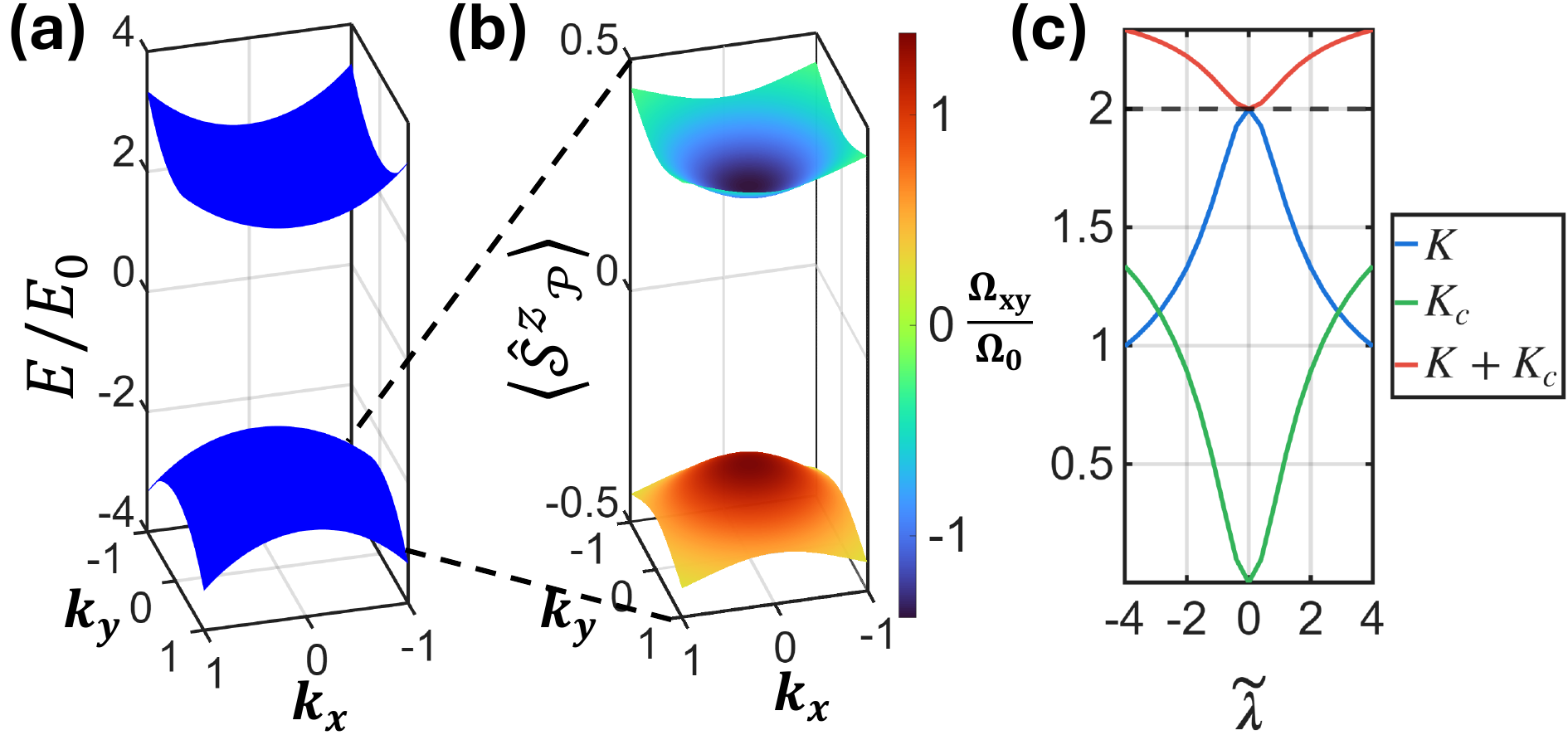}
  \caption{  Schematics of (a) the band structure and (b) the spectrum of $\mathcal{\hat{S}}^z_{\,\,\mathcal{P}}$ for the SCI model in Eq.~\eqref{eq:model} with $\Delta = m$, $\tilde{\mu} = 1.2$, $\tilde{\lambda} = 1.6$, and $E_0=m\Delta^2/2$. Colormap shows the Berry curvature distribution $\Omega_{xy}$ with $\Omega_0 = 10^{-4}$. (c) $K$, $K_c$, and $K+K_c$ as functions of $\tilde{\lambda}$ with $\tilde{\mu}=1$ and $\Delta=m$. Dashed line marks the topological bound set by $\sum_{\alpha=\pm}|C_\alpha|$.}
  \label{fig:model}
\end{figure}

\paragraph*{\textbf{Quantum geometry in a spin-$U(1)$ symmetry-breaking SCI}---}To illustrate how symmetry breaking influences quantum geometry, we consider an SCI model to which we add a spin-$U(1)$ symmetry-breaking SOC near the band inversion point. Note that the SOC can enhance band mixing and thus increase the quantum weight, while spin-$U(1)$ symmetry breaking relaxes the associated bounds. To examine this interplay, we consider the Hamiltonian:
\begin{align}\label{eq:model}
    H = \mathbf{d}(\vec{k})\cdot\mathbf{\Gamma}.
\end{align}
Here, $\mathbf{\Gamma} = (\tau_x,\tau_y,\tau_z\sigma_z,\tau_z\sigma_x,\tau_z\sigma_y)$ are the Gamma matrices, where $\boldsymbol\tau$ and $\boldsymbol\sigma$ refer to the Pauli matrices for orbital and spin degrees of freedom, respectively, and they satisfy $\{\Gamma_i,\Gamma_j\}=2\delta_{ij}$ and $[\Gamma_i,\Gamma_j]=2i\epsilon_{ij}^{\,\,\,\, k}\Gamma_k$. $\mathbf{d}(\vec{k})$ is a vector given by
\begin{align}\label{eq:d_k}
  \mathbf{d}(\vec{k}) = (\Delta k_x, \Delta k_y, \frac{k^2}{2m}-\mu,\lambda,0),
\end{align}
in which $\lambda$ controls the strength of the SOC. We set $\hbar=1$ for simplicity. The dispersions are given by $\epsilon_{\pm}(\vec{k})=\pm\|\mathbf{d}(\vec{k})\| = \pm\sqrt{\Delta^2k^2+(\frac{k^2}{2m}-\mu)^2+\lambda^2}$. Each band is doubly degenerate due to the PT symmetry, where $P=\tau_x$ is the inversion symmetry and $T=i\sigma_y\mathcal{K}$ is the time-reversal symmetry with $\mathcal{K}$ the complex conjugation operator. A schematic of the band structure is shown in Fig.~\ref{fig:model} (a). 

\par For $\lambda = 0$, Eq.~\eqref{eq:model} preserves spin-$U(1)$ symmetry and describes a time-reversal-symmetry-breaking SCI when $\mu>0$, where $K\geq\sum_{\alpha=\uparrow,\downarrow}|C_\alpha|=2$. The cases of $\mu=0$ and $\mu<0$ describe the critical and the topologically trivial phases, respectively. For $\lambda\neq0$, Eq.~\eqref{eq:model} breaks spin-$U(1)$ symmetry and opens a band gap when $\mu=0$. In this case, the band topology of the occupied states can only be captured by the projected spectrum of $\mathcal{\hat{S}}^z_{\,\,\mathcal{P}}$ with $P$ being the projection operator onto the occupied states. The projected spectrum of $\mathcal{\hat{S}}^z_{\,\,\mathcal{P}}$ hosts Chern numbers $C_\pm = \mp 1$ for sectors with positive/negative eigenvalues when $\mu>0$ (Fig.~\ref{fig:model}(b)).

Figure~\ref{fig:model}(c) shows the $K$, $K_c$, and $K+K_c$ computed as functions of $\tilde{\lambda}$ with $\tilde{\mu}=1$, where $\tilde{\mu}=\mu/E_0$, $\tilde{\lambda}=\lambda/E_0$, and $E_0=m\Delta^2/2$. As $\tilde{\lambda}$ increases, $K$ decreases while $K_c$ increases. Notably, $K < \sum_{\alpha=\pm}|C_\alpha| = 2$ for $\tilde{\lambda}\neq0$, in sharp contrast to the spin-$U(1)$-symmetric phase where the conventional topological bound requires $K\geq\sum_{\alpha=\uparrow,\downarrow}|C_\alpha|=2$. This behavior can be understood from the quantum metric of the occupied states. Using the projection operator $P(\vec{k})=(1-\mathbf{d}(\vec{k})\cdot\mathbf{\Gamma})/2$, the quantum metric takes the form $g^{\mu\nu}(\vec{k})=(\partial^\mu\mathbf{d}(\vec{k})\cdot\partial^\nu\mathbf{d}(\vec{k}))/2$, yielding
\begin{equation}\label{eq:g_model}    
\text{tr}[g(\vec{k})] = \frac{\Delta^2}{\|\mathbf{d}(\vec{k})\|^2} + \frac{k^2E_0^2}{2m^2\|\mathbf{d}(\vec{k})\|^4} \left[\tilde{\lambda}^2 + 4(\tilde{\mu} - 1) \right]^2.
\end{equation}
Increasing $\lambda$ enlarges $\|\mathbf{d}(\vec{k})\|$ according to Eq.~\eqref{eq:d_k}, which suppresses the quantum metric via Eq.~\eqref{eq:g_model} and consequently reduces $K$ below the topological bound set by the Chern number of each sector in the projected spectrum. In contrast, the bound of Eq.~\eqref{eq:result}, $K+K_c\geq\sum_{\alpha=\pm}|C_\alpha| = 2$, is always satisfied, demonstrating its applicability beyond SPT phases.

\paragraph*{\textbf{Conclusions}---} We discuss effects of symmetry-breaking perturbations on quantum geometry in topological systems at a quantitative level and show how the concept of topological bounds can be extended beyond the SPT phases as embodied in Eq.~\eqref{eq:result}. Although we have focused on the SCIs, Eq.~\eqref{eq:result} is more general and applicable to systems beyond the SCIs through Eq.~\eqref{eq:project_o}, such as the mirror Chern insulators \cite{Liu2014, Hsieh2012}, orbital Hall effect materials \cite{PhysRevB.111.195102, PhysRevLett.126.056601}, and tailored lattice models \cite{wang2023feature}. Our study provides insight into the subtle interplay between topology, quantum geometry, and symmetry breaking underlying physical responses of quantum materials, and identifies new pathways for tailoring their electronic structures in more realistic, symmetry-broken settings. 
\paragraph*{\textbf{Acknowledgements}---} The work at Northeastern University was supported by the National Science Foundation through the Expand-QISE award NSF-OMA-2329067 and benefited from the resources of Northeastern University’s Advanced Scientific Computation Center, the Explorer Cluster, the Massachusetts Technology Collaborative award MTC-22032, and the Quantum Materials and Sensing Institute. The work at MIT was supported by the Air
Force Office of Scientific Research under award number FA2386-24-1-4043. YO was supported in part by Grant No. NSF PHY-1748958 to the Kavli Institute for Theoretical Physics (KITP), the Heising-Simons Foundation, and the Simons Foundation (216179, LB). H.L. acknowledges the support by the National Science and Technology Council (NSTC) in Taiwan under grant number NSTC 114-2112-M-001-055-MY3. The Global Seed Funds provided by MIT International Science and Technology Initiatives (MISTI) under the program of MIT Greater China Fund for Innovation are acknowledged.

\paragraph*{\textbf{Data availability}---}The data that support the findings of this article are available from the authors upon reasonable request.

\bibliography{ref_aps}

@misc{SM,
note={see Supplemental Materials at [link-inserted-by-the-editor] for more details, which includes Ref.~\cite{PhysRevR.7.L042011}.}
}

@misc{footnote,
note={ An interesting attempt to address this problem has been recently made in systems with spin Chern number $C_s = 2$ \cite{PhysRevR.7.L042011}. However, as detailed in Supplemental Materials (SM) \cite{SM}, our analysis indicates a subtle inconsistency in the proof of Ref. \cite{PhysRevR.7.L042011}.}
}

@article{Resta_2002,
doi = {10.1088/0953-8984/14/20/201},
url = {https://doi.org/10.1088/0953-8984/14/20/201},
year = {2002},
month = {may},
publisher = {},
volume = {14},
number = {20},
pages = {R625},
author = {Raffaele Resta},
title = {Why are insulators  insulating
and metals conducting?},
journal = {Journal of Physics: Condensed Matter}
}

@article{PhysRevX.14.011052,
  title = {Fundamental Bound on Topological Gap},
  author = {Onishi, Yugo and Fu, Liang},
  journal = {Phys. Rev. X},
  volume = {14},
  issue = {1},
  pages = {011052},
  numpages = {12},
  year = {2024},
  month = {Mar},
  publisher = {American Physical Society},
  doi = {10.1103/PhysRevX.14.011052},
  url = {https://link.aps.org/doi/10.1103/PhysRevX.14.011052}
}

@article{PhysRevLett.133.206602,
  title = {Topological Bound on the Structure Factor},
  author = {Onishi, Yugo and Fu, Liang},
  journal = {Phys. Rev. Lett.},
  volume = {133},
  issue = {20},
  pages = {206602},
  numpages = {6},
  year = {2024},
  month = {Nov},
  publisher = {American Physical Society},
  doi = {10.1103/PhysRevLett.133.206602},
  url = {https://link.aps.org/doi/10.1103/PhysRevLett.133.206602}
}

@article{PhysRevB.112.035158,
  title = {Geometric bound on the structure factor and a harmonic condition on band geometry},
  author = {Onishi, Yugo and Avdoshkin, Alexander and Fu, Liang},
  journal = {Phys. Rev. B},
  volume = {112},
  issue = {3},
  pages = {035158},
  numpages = {6},
  year = {2025},
  month = {Jul},
  publisher = {American Physical Society},
  doi = {10.1103/8ng1-bwf6},
  url = {https://link.aps.org/doi/10.1103/8ng1-bwf6}
}

@Article{Yu2025,
author={Yu, Jiabin
and Bernevig, B. Andrei
and Queiroz, Raquel
and Rossi, Enrico
and T{\"o}rm{\"a}, P{\"a}ivi
and Yang, Bohm-Jung},
title={Quantum geometry in quantum materials},
journal={npj Quantum Materials},
year={2025},
month={Oct},
day={10},
volume={10},
number={1},
pages={101},
issn={2397-4648},
doi={10.1038/s41535-025-00801-3},
url={https://doi.org/10.1038/s41535-025-00801-3}
}

@article{PhysRevResearch.7.023158,
  title = {Quantum weight: A fundamental property of quantum many-body systems},
  author = {Onishi, Yugo and Fu, Liang},
  journal = {Phys. Rev. Res.},
  volume = {7},
  issue = {2},
  pages = {023158},
  numpages = {11},
  year = {2025},
  month = {May},
  publisher = {American Physical Society},
  doi = {10.1103/PhysRevResearch.7.023158},
  url = {https://link.aps.org/doi/10.1103/PhysRevResearch.7.023158}
}

@article{PhysRevB.90.165139,
  title = {Band geometry of fractional topological insulators},
  author = {Roy, Rahul},
  journal = {Phys. Rev. B},
  volume = {90},
  issue = {16},
  pages = {165139},
  numpages = {7},
  year = {2014},
  month = {Oct},
  publisher = {American Physical Society},
  doi = {10.1103/PhysRevB.90.165139},
  url = {https://link.aps.org/doi/10.1103/PhysRevB.90.165139}
}

@article{PhysRevLett.135.086401,
  title = {Universal Wilson Loop Bound of Quantum Geometry},
  author = {Yu, Jiabin and Herzog-Arbeitman, Jonah and Bernevig, B. Andrei},
  journal = {Phys. Rev. Lett.},
  volume = {135},
  issue = {8},
  pages = {086401},
  numpages = {8},
  year = {2025},
  month = {Aug},
  publisher = {American Physical Society},
  doi = {10.1103/mp2c-zzkt},
  url = {https://link.aps.org/doi/10.1103/mp2c-zzkt}
}

@misc{yu2025wilsonloop,
      title={Wilson-Loop-Ideal Bands and General Idealization}, 
      author={Jiabin Yu and Biao Lian and Shinsei Ryu},
      year={2025},
      eprint={2509.05410},
      archivePrefix={arXiv},
      primaryClass={cond-mat.mes-hall},
      url={https://arxiv.org/abs/2509.05410}, 
}

@article{PhysRevR.7.L042011,
  title = {Quantum geometric bounds in spinful systems with trivial band topology},
  author = {Jankowski, Wojciech J. and Slager, Robert-Jan and Lange, Gunnar F.},
  journal = {Phys. Rev. Res.},
  volume = {7},
  issue = {4},
  pages = {L042011},
  numpages = {8},
  year = {2025},
  month = {Oct},
  publisher = {American Physical Society},
  doi = {10.1103/zlxq-fxgc},
  url = {https://link.aps.org/doi/10.1103/zlxq-fxgc}
}

@article{PhysRevB.95.024515,
  title = {Band geometry, Berry curvature, and superfluid weight},
  author = {Liang, Long and Vanhala, Tuomas I. and Peotta, Sebastiano and Siro, Topi and Harju, Ari and T\"orm\"a, P\"aivi},
  journal = {Phys. Rev. B},
  volume = {95},
  issue = {2},
  pages = {024515},
  numpages = {16},
  year = {2017},
  month = {Jan},
  publisher = {American Physical Society},
  doi = {10.1103/PhysRevB.95.024515},
  url = {https://link.aps.org/doi/10.1103/PhysRevB.95.024515}
}

@article{Peotta2015,
author={Peotta, Sebastiano
and T{\"o}rm{\"a}, P{\"a}ivi},
title={Superfluidity in topologically nontrivial flat bands},
journal={Nature Communications},
year={2015},
month={Nov},
day={20},
volume={6},
number={1},
pages={8944},
issn={2041-1723},
doi={10.1038/ncomms9944},
url={https://doi.org/10.1038/ncomms9944}
}

@article{PhysRevLett.124.167002,
  title = {Topology-Bounded Superfluid Weight in Twisted Bilayer Graphene},
  author = {Xie, Fang and Song, Zhida and Lian, Biao and Bernevig, B. Andrei},
  journal = {Phys. Rev. Lett.},
  volume = {124},
  issue = {16},
  pages = {167002},
  numpages = {6},
  year = {2020},
  month = {Apr},
  publisher = {American Physical Society},
  doi = {10.1103/PhysRevLett.124.167002},
  url = {https://link.aps.org/doi/10.1103/PhysRevLett.124.167002}
}

@article{PhysRevLett.128.087002,
  title = {Superfluid Weight Bounds from Symmetry and Quantum Geometry in Flat Bands},
  author = {Herzog-Arbeitman, Jonah and Peri, Valerio and Schindler, Frank and Huber, Sebastian D. and Bernevig, B. Andrei},
  journal = {Phys. Rev. Lett.},
  volume = {128},
  issue = {8},
  pages = {087002},
  numpages = {8},
  year = {2022},
  month = {Feb},
  publisher = {American Physical Society},
  doi = {10.1103/PhysRevLett.128.087002},
  url = {https://link.aps.org/doi/10.1103/PhysRevLett.128.087002}
}

@article{PhysRevB.109.155143,
  title = {Feature-energy duality of topological boundary states in a multilayer quantum spin Hall insulator},
  author = {Yao, Yueh-Ting and Zhou, Xiaoting and Hung, Yi-Chun and Lin, Hsin and Bansil, Arun and Chang, Tay-Rong},
  journal = {Phys. Rev. B},
  volume = {109},
  issue = {15},
  pages = {155143},
  numpages = {7},
  year = {2024},
  month = {Apr},
  publisher = {American Physical Society},
  doi = {10.1103/PhysRevB.109.155143},
  url = {https://link.aps.org/doi/10.1103/PhysRevB.109.155143}
}

@article{PhysRevB.111.195102,
  title = {Topological characteristics and bulk-boundary correspondence in the orbital Hall effect},
  author = {Wang, Baokai and Hung, Yi-Chun and Lin, Hsin and Li, Sheng and He, Rui-Hua and Bansil, Arun},
  journal = {Phys. Rev. B},
  volume = {111},
  issue = {19},
  pages = {195102},
  numpages = {6},
  year = {2025},
  month = {May},
  publisher = {American Physical Society},
  doi = {10.1103/PhysRevB.111.195102},
  url = {https://link.aps.org/doi/10.1103/PhysRevB.111.195102}
}

@misc{wang2023feature,
      title={Feature Spectrum Topology}, 
      author={Baokai Wang and Yi-Chun Hung and Xiaoting Zhou and Tzen Ong and Hsin Lin},
      year={2023},
      eprint={2310.14832},
      archivePrefix={arXiv},
      primaryClass={cond-mat.mtrl-sci},
      url={https://arxiv.org/abs/2310.14832}, 
}

@article{Yao_2026,
doi = {10.1088/1361-6633/ae2a68},
url = {https://doi.org/10.1088/1361-6633/ae2a68},
year = {2025},
month = {dec},
publisher = {IOP Publishing},
volume = {89},
number = {1},
pages = {018001},
author = {Yao, Yueh-Ting and Chu, Chia-Hung and Bansil, Arun and Lin, Hsin and Chang, Tay-Rong},
title = {Orbital topology induced orbital Hall effect in two-dimensional insulators},
journal = {Reports on Progress in Physics}
}

@Article{Lin2024,
author={Lin, Kuan-Sen
and Palumbo, Giandomenico
and Guo, Zhaopeng
and Hwang, Yoonseok
and Blackburn, Jeremy
and Shoemaker, Daniel P.
and Mahmood, Fahad
and Wang, Zhijun
and Fiete, Gregory A.
and Wieder, Benjamin J.
and Bradlyn, Barry},
title={Spin-resolved topology and partial axion angles in three-dimensional insulators},
journal={Nature Communications},
year={2024},
month={Jan},
day={16},
volume={15},
number={1},
pages={550},
issn={2041-1723},
doi={10.1038/s41467-024-44762-w},
url={https://doi.org/10.1038/s41467-024-44762-w}
}

@misc{shulman2010robust,
      title={Robust extended states in a topological bulk model with even spin-Chern invariant}, 
      author={Hadassah Shulman and Emil Prodan},
      year={2010},
      eprint={1011.5456},
      archivePrefix={arXiv},
      primaryClass={cond-mat.dis-nn}
}

@article{PhysRevB.80.125327,
  title = {Robustness of the spin-Chern number},
  author = {Prodan, Emil},
  journal = {Phys. Rev. B},
  volume = {80},
  issue = {12},
  pages = {125327},
  numpages = {7},
  year = {2009},
  month = {Sep},
  publisher = {American Physical Society},
  doi = {10.1103/PhysRevB.80.125327},
  url = {https://link.aps.org/doi/10.1103/PhysRevB.80.125327}
}

@article{PhysRevLett.126.056601,
  title = {Disentangling Orbital and Valley Hall Effects in Bilayers of Transition Metal Dichalcogenides},
  author = {Cysne, Tarik P. and Costa, Marcio and Canonico, Luis M. and Nardelli, M. Buongiorno and Muniz, R. B. and Rappoport, Tatiana G.},
  journal = {Phys. Rev. Lett.},
  volume = {126},
  issue = {5},
  pages = {056601},
  numpages = {7},
  year = {2021},
  month = {Feb},
  publisher = {American Physical Society},
  doi = {10.1103/PhysRevLett.126.056601},
  url = {https://link.aps.org/doi/10.1103/PhysRevLett.126.056601}
}

@article{RevModPhys.88.021004,
  title = {Colloquium: Topological band theory},
  author = {Bansil, A. and Lin, Hsin and Das, Tanmoy},
  journal = {Rev. Mod. Phys.},
  volume = {88},
  issue = {2},
  pages = {021004},
  numpages = {37},
  year = {2016},
  month = {Jun},
  publisher = {American Physical Society},
  doi = {10.1103/RevModPhys.88.021004},
  url = {https://link.aps.org/doi/10.1103/RevModPhys.88.021004}
}

@article{PhysRevB.78.195424,
  title = {Topological field theory of time-reversal invariant insulators},
  author = {Qi, Xiao-Liang and Hughes, Taylor L. and Zhang, Shou-Cheng},
  journal = {Phys. Rev. B},
  volume = {78},
  issue = {19},
  pages = {195424},
  numpages = {43},
  year = {2008},
  month = {Nov},
  publisher = {American Physical Society},
  doi = {10.1103/PhysRevB.78.195424},
  url = {https://link.aps.org/doi/10.1103/PhysRevB.78.195424}
}

@article{PhysRevB.76.045302,
  title = {Topological insulators with inversion symmetry},
  author = {Fu, Liang and Kane, C. L.},
  journal = {Phys. Rev. B},
  volume = {76},
  issue = {4},
  pages = {045302},
  numpages = {17},
  year = {2007},
  month = {Jul},
  publisher = {American Physical Society},
  doi = {10.1103/PhysRevB.76.045302},
  url = {https://link.aps.org/doi/10.1103/PhysRevB.76.045302}
}

@article{PhysRevB.75.121306,
  title = {Topological invariants of time-reversal-invariant band structures},
  author = {Moore, J. E. and Balents, L.},
  journal = {Phys. Rev. B},
  volume = {75},
  issue = {12},
  pages = {121306},
  numpages = {4},
  year = {2007},
  month = {Mar},
  publisher = {American Physical Society},
  doi = {10.1103/PhysRevB.75.121306},
  url = {https://link.aps.org/doi/10.1103/PhysRevB.75.121306}
}

@article{PhysRevB.74.195312,
  title = {Time reversal polarization and a ${Z}_{2}$ adiabatic spin pump},
  author = {Fu, Liang and Kane, C. L.},
  journal = {Phys. Rev. B},
  volume = {74},
  issue = {19},
  pages = {195312},
  numpages = {13},
  year = {2006},
  month = {Nov},
  publisher = {American Physical Society},
  doi = {10.1103/PhysRevB.74.195312},
  url = {https://link.aps.org/doi/10.1103/PhysRevB.74.195312}
}

@Article{Po2017,
author={Po, Hoi Chun
and Vishwanath, Ashvin
and Watanabe, Haruki},
title={Symmetry-based indicators of band topology in the 230 space groups},
journal={Nature Communications},
year={2017},
month={Jun},
day={30},
volume={8},
number={1},
pages={50},
issn={2041-1723},
doi={10.1038/s41467-017-00133-2},
url={https://doi.org/10.1038/s41467-017-00133-2}
}

@ARTICLE{Bradlyn2017-pj,
  title    = "Topological quantum chemistry",
  author   = "Bradlyn, Barry and Elcoro, L and Cano, Jennifer and Vergniory, M
              G and Wang, Zhijun and Felser, C and Aroyo, M I and Bernevig, B
              Andrei",
  journal  = "Nature",
  volume   =  547,
  number   =  7663,
  pages    = "298--305",
  month    =  jul,
  year     =  2017
}

@article{PhysRevX.7.041069,
  title = {Topological Classification of Crystalline Insulators through Band Structure Combinatorics},
  author = {Kruthoff, Jorrit and de Boer, Jan and van Wezel, Jasper and Kane, Charles L. and Slager, Robert-Jan},
  journal = {Phys. Rev. X},
  volume = {7},
  issue = {4},
  pages = {041069},
  numpages = {23},
  year = {2017},
  month = {Dec},
  publisher = {American Physical Society},
  doi = {10.1103/PhysRevX.7.041069},
  url = {https://link.aps.org/doi/10.1103/PhysRevX.7.041069}
}

@article{PhysRevLett.108.196806,
  title = {Connection of Edge States to Bulk Topological Invariance in a Quantum Spin Hall State},
  author = {Li, Huichao and Sheng, L. and Xing, D. Y.},
  journal = {Phys. Rev. Lett.},
  volume = {108},
  issue = {19},
  pages = {196806},
  numpages = {5},
  year = {2012},
  month = {May},
  publisher = {American Physical Society},
  doi = {10.1103/PhysRevLett.108.196806},
  url = {https://link.aps.org/doi/10.1103/PhysRevLett.108.196806}
}

@article{YANG2018723,
title = {Topological phase transition in anisotropic square-octagon lattice with spin–orbit coupling and exchange field},
journal = {Physics Letters A},
volume = {382},
number = {10},
pages = {723-728},
year = {2018},
issn = {0375-9601},
doi = {https://doi.org/10.1016/j.physleta.2017.12.051},
url = {https://www.sciencedirect.com/science/article/pii/S0375960118300033},
author = {Yuan Yang and Jian Yang and Xiaobing Li and Yue Zhao}
}

@Article{Hsieh2012,
author={Hsieh, Timothy H.
and Lin, Hsin
and Liu, Junwei
and Duan, Wenhui
and Bansil, Arun
and Fu, Liang},
title={Topological crystalline insulators in the SnTe material class},
journal={Nature Communications},
year={2012},
month={Jul},
day={31},
volume={3},
number={1},
pages={982},
issn={2041-1723},
doi={10.1038/ncomms1969},
url={https://doi.org/10.1038/ncomms1969}
}

@Article{Liu2014,
author={Liu, Junwei
and Hsieh, Timothy H.
and Wei, Peng
and Duan, Wenhui
and Moodera, Jagadeesh
and Fu, Liang},
title={Spin-filtered edge states with an electrically tunable gap in a two-dimensional topological crystalline insulator},
journal={Nature Materials},
year={2014},
month={Feb},
day={01},
volume={13},
number={2},
pages={178-183},
issn={1476-4660},
doi={10.1038/nmat3828},
url={https://doi.org/10.1038/nmat3828}
}

@article{
doi:10.1126/science.adf1506,
author = {Anyuan Gao  and Yu-Fei Liu  and Jian-Xiang Qiu  and Barun Ghosh  and Thaís V. Trevisan  and Yugo Onishi  and Chaowei Hu  and Tiema Qian  and Hung-Ju Tien  and Shao-Wen Chen  and Mengqi Huang  and Damien Bérubé  and Houchen Li  and Christian Tzschaschel  and Thao Dinh  and Zhe Sun  and Sheng-Chin Ho  and Shang-Wei Lien  and Bahadur Singh  and Kenji Watanabe  and Takashi Taniguchi  and David C. Bell  and Hsin Lin  and Tay-Rong Chang  and Chunhui Rita Du  and Arun Bansil  and Liang Fu  and Ni Ni  and Peter P. Orth  and Qiong Ma  and Su-Yang Xu },
title = {Quantum metric nonlinear Hall effect in a topological antiferromagnetic heterostructure},
journal = {Science},
volume = {381},
number = {6654},
pages = {181-186},
year = {2023},
doi = {10.1126/science.adf1506},
URL = {https://www.science.org/doi/abs/10.1126/science.adf1506}}

@article{
doi:10.1126/sciadv.ado1761,
author = {Barun Ghosh  and Yugo Onishi  and Su-Yang Xu  and Hsin Lin  and Liang Fu  and Arun Bansil },
title = {Probing quantum geometry through optical conductivity and magnetic circular dichroism},
journal = {Science Advances},
volume = {10},
number = {51},
pages = {eado1761},
year = {2024},
doi = {10.1126/sciadv.ado1761},
URL = {https://www.science.org/doi/abs/10.1126/sciadv.ado1761}}

@Article{Wang2023,
author={Wang, Naizhou
and Kaplan, Daniel
and Zhang, Zhaowei
and Holder, Tobias
and Cao, Ning
and Wang, Aifeng
and Zhou, Xiaoyuan
and Zhou, Feifei
and Jiang, Zhengzhi
and Zhang, Chusheng
and Ru, Shihao
and Cai, Hongbing
and Watanabe, Kenji
and Taniguchi, Takashi
and Yan, Binghai
and Gao, Weibo},
title={Quantum-metric-induced nonlinear transport in a topological antiferromagnet},
journal={Nature},
year={2023},
month={Sep},
day={01},
volume={621},
number={7979},
pages={487-492},
issn={1476-4687},
doi={10.1038/s41586-023-06363-3},
url={https://doi.org/10.1038/s41586-023-06363-3}
}

@article{PhysRevB.56.12847,
  title = {Maximally localized generalized Wannier functions for composite energy bands},
  author = {Marzari, Nicola and Vanderbilt, David},
  journal = {Phys. Rev. B},
  volume = {56},
  issue = {20},
  pages = {12847--12865},
  numpages = {0},
  year = {1997},
  month = {Nov},
  publisher = {American Physical Society},
  doi = {10.1103/PhysRevB.56.12847},
  url = {https://link.aps.org/doi/10.1103/PhysRevB.56.12847}
}

@article{PhysRevB.112.155158,
  title = {Quantum geometric bounds for observables: Linear responses, Drude weight, and orbital magnetization},
  author = {Shinada, Koki and Nagaosa, Naoto},
  journal = {Phys. Rev. B},
  volume = {112},
  issue = {15},
  pages = {155158},
  numpages = {12},
  year = {2025},
  month = {Oct},
  publisher = {American Physical Society},
  doi = {10.1103/qxbl-qd4f},
  url = {https://link.aps.org/doi/10.1103/qxbl-qd4f}
}

@book{Horn_Johnson_1985, place={Cambridge}, title={Matrix Analysis}, publisher={Cambridge University Press}, author={Horn, Roger A. and Johnson, Charles R.}, year={1985}}

@book{conway1994course,
  title={A Course in Functional Analysis},
  author={Conway, J.B.},
  isbn={9780387972459},
  lccn={97122669},
  series={Graduate Texts in Mathematics},
  year={1994},
  publisher={Springer New York}
}

@book{fubini1904sulle,
  title={Sulle metriche definite da una forma hermitiana: nota},
  author={Fubini, Guido},
  year={1904},
  publisher={Office graf. C. Ferrari}
}

@Article{Study1905,
author={Study, E.},
title={K{\"u}rzeste Wege im komplexen Gebiet},
journal={Mathematische Annalen},
year={1905},
month={Sep},
day={01},
volume={60},
number={3},
pages={321-378},
issn={1432-1807},
doi={10.1007/BF01457616},
url={https://doi.org/10.1007/BF01457616}
}

@article{provost1980,
  title = {Riemannian Structure on Manifolds of Quantum States},
  author = {Provost, J. P. and Vallee, G.},
  year = 1980,
  month = sep,
  journal = {Communications in Mathematical Physics},
  volume = {76},
  number = {3},
  pages = {289--301},
  issn = {1432-0916},
  doi = {10.1007/BF02193559},
  urldate = {2023-12-06}
}

@article{PhysRevB.112.075116,
  title = {Low-energy optical absorption in correlated insulators: Projected sum rules and the role of quantum geometry},
  author = {Mao, Dan and Mendez-Valderrama, Juan Felipe and Chowdhury, Debanjan},
  journal = {Phys. Rev. B},
  volume = {112},
  issue = {7},
  pages = {075116},
  numpages = {18},
  year = {2025},
  month = {Aug},
  publisher = {American Physical Society},
  doi = {10.1103/xmz7-jgl6},
  url = {https://link.aps.org/doi/10.1103/xmz7-jgl6}
}

\setcounter{equation}{0}
\setcounter{figure}{0}
\setcounter{table}{0}

\renewcommand{\theequation}{S\arabic{equation}}
\renewcommand{\thefigure}{S\arabic{figure}}
\renewcommand{\thetable}{S\arabic{table}}
\renewcommand{\bibnumfmt}[1]{[S#1]}
\renewcommand{\citenumfont}[1]{S#1}
\newcommand{\bk}{\boldsymbol\kappa}

\newcommand{\beginsupplement}{%
  \setcounter{equation}{0}
  \renewcommand{\theequation}{S\arabic{equation}}%
  \setcounter{table}{0}
  \renewcommand{\thetable}{S\arabic{table}}%
  \setcounter{figure}{0}
  \renewcommand{\thefigure}{S\arabic{figure}}%
  \setcounter{section}{0}
  \renewcommand{\thesection}{S\Roman{section}}%
  \setcounter{subsection}{0}
  \renewcommand{\thesubsection}{S\Roman{section}.\Alph{subsection}}%
}

\clearpage
\pagebreak
\widetext
\begin{center}
\textbf{\large Supplemental Material: Extending Topological Bound on Quantum Weight Beyond Symmetry-Protected Topological Phases}
\end{center}
\tableofcontents

\section{S1. Remarks on the proof of the band-gap bound for SCIs with $C_s=2$}
\par Here, we reexamine some nuances related to the discussion in  Ref.~\cite{PhysRevR.7.L042011} of the band-gap bound in SCIs with spin Chern number $C_s=2$.  We note that although Ref.~\cite{PhysRevR.7.L042011} also considered the spectrum of the projected spin operator $\mathcal{\hat{S}}^z_{\,\,\mathcal{P}}=P \hat{S}^z P$ and obtained a result similar to our Eq.~(6) of the main text (see Eq.~(3) of the Supplemental Material (SM) of Ref. \cite{PhysRevR.7.L042011}); however, an important aspect appears to have been left unexplored. In the formulation of Ref.~\cite{PhysRevR.7.L042011}, the eigenstates of $\mathcal{\hat{S}}^z_{\,\,\mathcal{P}}$ are expressed as linear combinations of the energy eigenstates within the occupied Hilbert subspace. However, the $\vec{k}$-dependence of the coefficients in this linear combination ($C_{nm}^{\sigma}$ in their notation and $U_{mn}$ in our notation) was neglected. In general, these coefficients depend on $\vec{k}$ since the matrix elements of $\mathcal{\hat{S}}^z_{\,\,\mathcal{P}}$ vary with $\vec{k}$, which is also evident from the dispersive nature of the projected spectrum bands $\lambda_n(\vec{k})$. This $\vec{k}$-dependence becomes important when deriving the inequality that connects the quantum weight to the topological invariants because it leads to an additional term in the derivative of the wave function:
\begin{equation}
    \ket{\partial^\mu \tilde{u}_m^{(\alpha)}} = \sum_{n}(\partial^\mu U_{mn}) \ket{u_n} + U_{mn} \ket{\partial^\mu u_n}.
\end{equation}
With this correction, the second equality in Eq.~(19) of the SM of Ref. \cite{PhysRevR.7.L042011} will read (in their notation):
\begin{equation}\label{eq:correct}
    \sum_{n}^{N_{\text{occ}}} \left| \braket{ u_{n \mathbf{k}} | \partial_{k_i} u_{l \mathbf{k}}^{\sigma'} } \right|^2 = \sum_{n}^{N_{\text{occ}}} \left| \sum_{n'}^{N_{\text{occ}}} C_{n'l}^{\sigma'} \braket{ u_{n \mathbf{k}} | \partial_{k_i} u_{n' \mathbf{k}} } + (\partial_{k_i} C_{n'l}^{\sigma'}) \delta_{nn'} \right|^2.
\end{equation}
The second term on the right-hand side of Eq.~\eqref{eq:correct} modifies the last inequality in Eq.~(19) of the SM of Ref. \cite{PhysRevR.7.L042011}, which is central to their proof because it connects Eqs.~(23) and (24) of their SM. As a result, Eq.~(28) of the SM of Ref. \cite{PhysRevR.7.L042011} and, equivalently, their Eq.~(6) holds only in the special case where the coefficients connecting the energy eigenstates to the eigenstates of the projected spin spectrum are constant across the BZ: in fact, our SCI model with sufficiently strong SOC provides a counterexample.

\section{S2. Positive semidefiniteness and real symmetry of the quantum geometric tensors}
\par Positive semidefiniteness of $G^{\mu\nu}_\alpha$ and $G_{c,\alpha}^{\mu\nu}$ can be straightforwardly shown by reformulating them into $G^{\mu\nu}_\alpha=\text{Tr}[(Q_\alpha\partial^\mu P_\alpha)^\dagger(Q_\alpha\partial^\nu P_\alpha)]$ and $G^{\mu\nu}_{c,\alpha}=\text{Tr}[\left((P-P_\alpha)\partial^\mu P_\alpha\right)^\dagger\left((P-P_\alpha)\partial^\nu P_\alpha\right)]$. To prove that $G^{\mu\nu}_{c}$ is real-symmetric, note first that $P-P_\alpha=\sum_{\beta\neq\alpha}P_\beta$, and $G_c$ is Hermitian:
\begin{align}
    (G_{c}^{\mu\nu})^* & =(\sum_{\alpha\neq\beta}\sum_{n\in\alpha, m\in\beta}\braket{\partial^\mu \tilde{u}_n^{(\alpha)}|\tilde{u}_m^{(\beta)}}\braket{\tilde{u}_m^{(\beta)}|\partial^\nu\tilde{u}_n^{(\alpha)}})^* \notag
    \\ & = \sum_{\alpha\neq\beta}\sum_{n\in\alpha, m\in\beta}\braket{\partial^\nu \tilde{u}_n^{(\alpha)}|\tilde{u}_m^{(\beta)}}\braket{\tilde{u}_m^{(\beta)}|\partial^\mu\tilde{u}_n^{(\alpha)}} \notag
    \\ & = G_{c}^{\nu\mu}. \label{eq:G_c_real_1}
\end{align}
But, $G_c$ is symmetric:
\begin{align}
    G_{c}^{\mu\nu} & = \sum_{\alpha\neq\beta}\sum_{n\in\alpha, m\in\beta}(-\braket{\tilde{u}_n^{(\alpha)}|\partial^\mu\tilde{u}_m^{(\beta)}})(-\braket{\partial^\nu\tilde{u}_m^{(\beta)}|\tilde{u}_n^{(\alpha)}}) \notag
    \\ & = \sum_{\beta\neq\alpha}\sum_{n\in\beta, m\in\alpha}\braket{\partial^\nu\tilde{u}_m^{(\alpha)}|\tilde{u}_n^{(\beta)}}\braket{\tilde{u}_n^{(\beta)}|\partial^\mu\tilde{u}_m^{(\alpha)}} \notag
    \\ & = G_{c}^{\nu\mu}. \label{eq:G_c_real_2}
\end{align}
Combining Eqs.~\eqref{eq:G_c_real_1} and \eqref{eq:G_c_real_2} yields $G^{\mu\nu}_{c}=(G^{\mu\nu}_{c})^*$. Therefore, $G^{\mu\nu}_{c}$ is real-symmetric.

\section{S3. $K_c$ as a generic correction}
\par We show that $K_c$ provides a generic correction to $K$ in the topological bound defined in Eq.~\eqref{eq:result}, arising from $\hat{O}$\text{-}symmetry breaking, some rare exceptions notwithstanding. We first demonstrate that $K_c\neq0$ necessarily requires both $K\neq0$ and a broken $\hat{O}$\text{-}symmetry. We then show that additional constraints can still enforce $K_c=0$ in special cases, even when $\hat{O}$\text{-}symmetry is broken and $K$ remains nonzero. An illustrative example exhibiting this behavior is provided. 
\subsection{A. Necessary conditions for $K_c\neq0$: $K\neq0$ and broken $\hat{O}$-symmetry}
\par Here, we prove that (i) it is necessary to have $K\neq0$ and a broken $\hat{O}$-symmetry for $K_c\neq0$ and (ii) systems with $C_\alpha\neq0$ must have $K>0$.

For $\beta\neq\alpha$, we have
\begin{equation}    
    P_\beta\partial^\mu P_\alpha \propto P_\beta\partial^\mu \hat{\mathcal{O}}_{\mathcal{P}} P_\alpha.
\end{equation}
Since $\partial^{\mu}\mathcal{\hat{O}_{P}}=\partial^{\mu}P\hat{O}P + P\hat{O}\partial^{\mu}P$, $PP_\alpha=P_\alpha P=P_\alpha$, and $P_\alpha\partial^\mu PP=0$, we have
\begin{equation}\label{eq:equivalence}
    P_\beta\partial^\mu P_\alpha \propto P_\beta\partial^\mu PQ\hat{O}P_\alpha + P_\beta\hat{O}Q\partial^\mu PP_\alpha.
\end{equation}
Here, $P_\alpha\partial^{\mu}PP=0$ can be proven by noting, 
\begin{align}   
    P_\alpha\partial^{\mu}PP & = \sum_{n\in\alpha}\ket{\tilde{u}_n^{(\alpha)}}\bra{\tilde{u}_n^{(\alpha)}}\sum_{m\in\text{occ.}}\big(\ket{\partial^{\mu}u_m}\bra{u_m} + \ket{u_m}\bra{\partial^{\mu}u_m} \big)\sum_{l\in\text{occ.}}\ket{u_l}\bra{u_l}, \notag
    \\ & = \sum_{n\in\alpha}\sum_{m\in\text{occ.}}\ket{\tilde{u}_n^{(\alpha)}}\braket{\tilde{u}_n^{(\alpha)}|\partial^{\mu}u_m}\bra{u_m} - \sum_{n\in\alpha}\sum_{m\in\text{occ.}}\sum_{l\in\text{occ.}}\ket{\tilde{u}_n^{(\alpha)}}\braket{\tilde{u}_n^{(\alpha)}|u_m}\braket{u_m|\partial^{\mu}u_l}\bra{u_l}. \notag
    \\ & = \sum_{n\in\alpha}\sum_{m\in\text{occ.}}\ket{\tilde{u}_n^{(\alpha)}}\braket{\tilde{u}_n^{(\alpha)}|\partial^{\mu}u_m}\bra{u_m} - \sum_{n\in\alpha}\sum_{l\in\text{occ.}}\ket{\tilde{u}_n^{(\alpha)}}\braket{\tilde{u}_n^{(\alpha)}|P|\partial^{\mu}u_l}\bra{u_l} \notag
    \\ & = 0.
\end{align}
We can further simplify Eq.~\eqref{eq:equivalence} as:
\begin{equation}\label{eq:equivalence_result}
P_\beta\partial^\mu P_\alpha \propto \big( Q\partial^\mu P_\beta \big)^\dagger \big( Q\hat{O}P_\alpha \big) + \big( P_\beta\hat{O}Q \big)\big( Q\partial^\mu P_\alpha \big),
\end{equation}
where we have used $Q^2=Q$ and the identity $Q\partial^{\mu}PP_\beta= Q\partial^{\mu}P_\beta$, which can be proven straightforwardly as follows:
\begin{align}
 Q\partial^{\mu}PP_\beta & = \sum_{l\in\text{unocc.}}\sum_{m\in\text{occ.}}\sum_{n\in\beta}\ket{u_l}\braket{u_l|\partial^{\mu}u_m}\braket{u_m|\tilde{u}_n^{(\beta)}}\bra{\tilde{u}_n^{(\beta)}} \notag
 \\ & = - \sum_{l\in\text{unocc.}}\sum_{m\in\text{occ.}}\sum_{n\in\beta}\ket{u_l}\braket{\partial^{\mu}u_l|u_m}\braket{u_m|\tilde{u}_n^{(\beta)}}\bra{\tilde{u}_n^{(\beta)}} \notag
 \\ & = - \sum_{l\in\text{unocc.}}\sum_{n\in\beta}\ket{u_l}\braket{\partial^{\mu}u_l|P|\tilde{u}_n^{(\beta)}}\bra{\tilde{u}_n^{(\beta)}} \notag
 \\ & = - \sum_{l\in\text{unocc.}}\sum_{n\in\beta}\ket{u_l}\braket{\partial^{\mu}u_l|\tilde{u}_n^{(\beta)}}\bra{\tilde{u}_n^{(\beta)}} \notag
 \\ & = \sum_{l\in\text{unocc.}}\sum_{n\in\beta}\ket{u_l}\braket{u_l|\partial^{\mu}\tilde{u}_n^{(\beta)}}\bra{\tilde{u}_n^{(\beta)}} \notag
 \\ & = Q\partial^{\mu}P_\beta.
\end{align}
Combining $G^{\mu\nu}=\sum_{\alpha}\text{Tr}\left[\partial^{\mu}P_\alpha Q\partial^\nu P_\alpha\right]$ and Eq.~(16) of the main text with Eq.~\eqref{eq:equivalence_result} implies that if $K=0$, $K_c$ must also be zero. Therefore, it is necessary to have $K\neq0$ and a broken $\hat{O}$-symmetry for $K_c\neq0$.

\par In systems where $C_\alpha\neq0$ and $\hat{O}$-symmetry is broken, we have $Q_\alpha \partial^\mu P_\alpha\neq0$ since $g^{\mu\nu}_{\alpha}=\mathrm{Re}[G^{\mu\nu}_{\alpha}]$ and $\Omega^{\mu\nu}_{\alpha}=-2\mathrm{Im}[G^{\mu\nu}_{\alpha}]$, where $G_{\alpha}^{\mu\nu} = \text{Tr}[(\partial^\mu P_\alpha)Q_\alpha (\partial^\nu P_\alpha)]$ and $Q_\alpha=Q+\sum_{\beta\neq\alpha}P_\beta$. Therefore, if $Q\partial^{\mu}P_\alpha=0$, $\sum_{\beta\neq\alpha}P_\beta\partial^{\mu}P_\alpha$ must be nonzero. From Eq.~\eqref{eq:equivalence_result}, however, there must be at least one $\beta\neq\alpha$ satisfying $Q\partial^{\mu}P_\beta\neq0$. Therefore, it is impossible to have $Q\partial^{\mu}P_\alpha=0$ for all $\alpha$, which implies that $K$ must be greater than 0 in these systems. In other words, for systems with $C_\alpha\neq0$, if there are spectral gaps in the projected spectrum, $K>0$ no matter how strongly the symmetry is broken.
\subsection{B. An example of $K_c=0$ with a broken $\hat{O}$-symmetry and $K\neq0$}
\par We now discuss cases in which $K_c$ remains zero even if $K\neq0$ and the $\hat{O}$-symmetry is broken. Such cases appear in systems where an operator $\hat{\Xi}$ satisfies $[\hat{\Xi},\mathcal{\hat{O}}_{\mathcal{P}}]=0$, and the states in different sectors in the projected spectrum have different eigenvalues of $\hat{\Xi}$. Then, since $\mathcal{\hat{O}}_{\mathcal{P}}$ can be block-diagonalized by $\hat{\Xi}$, the projection operators to states in different sectors lie in different blocks. Therefore, $K_c=0$, because  $P_\beta\partial^{\mu}P_\alpha=0$ for all $\alpha\neq\beta$. The condition $[\hat{\Xi},\mathcal{\hat{O}}_{\mathcal{P}}]=0$ holds when $[\hat{\Xi},H]=[\hat{\Xi},\hat{O}]=0$ since:
    $\hat{\Xi}\mathcal{\hat{O}}_{\mathcal{P}}\hat{\Xi}^{-1} = \hat{\Xi} P\hat{O}P\hat{\Xi}^{-1} = P\hat{\Xi}\hat{O}\hat{\Xi}^{-1}P  = P\hat{O}P = \mathcal{\hat{O}}_{\mathcal{P}}.$
Note that the breaking of $\hat{O}$-symmetry requires $\hat{\Xi}$ to have degenerate eigenvalues.

\par $K_c$, however, remains a valuable descriptor for identifying the reduction of $K$ via $\hat{O}$-symmetry breaking, as the required condition is generally not expected to be satisfied in typical systems because it requires an additional symmetry $\hat{\Xi}$ with degenerate eigenvalues that commutes with $\hat{O}$, while assigning different eigenvalues of $\hat{\Xi}$ to states in different sectors in the projected spectrum.

\par We consider an illustrative model in which $K\neq0$ and the $\hat{O}$-symmetry is broken, but $K_c=0$ described by the Hamiltonian:
\begin{equation}\label{eq:counter_ex}
    H = \mu\tau_z+v(k_x\tau_x+k_y\tau_y)s_x+\lambda s_z,
\end{equation}
where $\boldsymbol\tau$ and $\boldsymbol s$ are the Pauli matrices for the orbital and spin degrees of freedom, respectively, and the basis can be written as $(\hat{c}_{a\uparrow},\hat{c}_{b\uparrow},\hat{c}_{a\downarrow},\hat{c}_{b\downarrow})$, where $a,b$ indicate the orbitals. The Hamiltonian of  Eq.~\eqref{eq:counter_ex} has the inversion symmetry $I=\tau_z$ and the additional symmetry $\hat{\Xi}=\tau_zs_z$. The spin-$U(1)$ symmetry is broken by the second term in Eq.~\eqref{eq:counter_ex}. 

\par In the case of $\hat{O}=\hat{S}^z$, since $[\hat{\Xi},\hat{O}]=0$, $\hat{\Xi}$ acts as a symmetry that can eliminate $K_c$, see discussion in the main text. To see this, we first block-diagonalize the Hamiltonian in Eq.~\eqref{eq:counter_ex} using the $\hat{\Xi}$ symmetry:
\begin{equation}\label{eq:H_new_basis}
    H = \begin{pmatrix} H_U & 0_{2\times2} \\ 0_{2\times2} & H_L \end{pmatrix}= \begin{pmatrix} \epsilon_U\hat{n}_U\cdot\vec{\tau} & 0_{2\times2} \\ 0_{2\times2} & \epsilon_L\hat{n}_L\cdot\vec{\tau} \end{pmatrix},
\end{equation}
where the basis becomes $(\hat{c}_{a\uparrow},\hat{c}_{b\downarrow},\hat{c}_{a\downarrow},\hat{c}_{b\uparrow})$. Here, $U$ and $L$ represent the upper and lower block and:
\begin{align}
    \epsilon_U & = \sqrt{(\mu+\lambda)^2+v^2k^2},
    \\ \epsilon_L & = \sqrt{(\mu-\lambda)^2+v^2k^2},
    \\ \hat{n}_U & = \frac{1}{\epsilon_U}(vk_x,vk_y,\mu+\lambda),
    \\ \hat{n}_L & = \frac{1}{\epsilon_L}(vk_x,vk_y,\mu-\lambda).
\end{align}
The quantum metric of the Hamiltonian in Eq.~\eqref{eq:counter_ex} thus equals
\begin{equation}
    g^{\mu\nu} = \frac{1}{2}(\partial^{\mu}\hat{n}_U\cdot\partial^{\nu}\hat{n}_U+\partial^{\mu}\hat{n}_L\cdot\partial^{\nu}\hat{n}_L).
\end{equation}
Through straightforward calculations, we obtain
\begin{equation}
    \text{tr}[g] = \frac{1}{2}\bigg( \frac{v^2}{(\mu+\lambda)^2+v^2k^2}\left[2-\frac{v^2k^2}{(\mu+\lambda)^2+v^2k^2} \right] + \frac{v^2}{(\mu-\lambda)^2+v^2k^2}\left[2-\frac{v^2k^2}{(\mu-\lambda)^2+v^2k^2} \right] \bigg),
\end{equation}
which does not equal $0$ everywhere in the $k$-space since $0< v^2/[(\mu\pm\lambda)^2+v^2k^2]<1/k^2$. Thus, $K\neq0$ in this model.

\par In the basis of $(\hat{c}_{a\uparrow},\hat{c}_{b\downarrow},\hat{c}_{a\downarrow},\hat{c}_{b\uparrow})$ used in Eq.~\eqref{eq:H_new_basis}, $\hat{S}^z=\tau_z\xi_z$, where $\boldsymbol\xi$ are Pauli matrices describing the block degree of freedom. Therefore, $\mathcal{\hat{S}}^z_{\,\,\mathcal{P}}$ equals
\begin{equation}\label{eq:counter_ex_psp}
\mathcal{\hat{S}}^z_{\,\,\mathcal{P}} = 
\begin{pmatrix}
    P_U\tau_zP_U & 0_{2\times2} \\ 0_{2\times2} & -P_L\tau_zP_L
\end{pmatrix}.
\end{equation}
Here, $P_b$ is the projection operator onto the occupied states, which equals
\begin{align}
    P_b & = \frac{1}{2}(\mathbf{1}-\hat{n}_b\cdot\vec{\tau}),
\end{align}
where $b=U,L$ describes the block degree of freedom. To calculate the eigenvalues and eigenvectors of $\mathcal{\hat{S}}^z_{\,\,\mathcal{P}}$ in Eq.~\eqref{eq:counter_ex_psp}, we consider the following parameterization:
\begin{equation}
    \hat{n}_{b} = (\sin\theta_b\cos\phi_b,\sin\theta_b\sin\phi_b,\cos\theta_b).
\end{equation}
Using $[\tau_i,\tau_j]=2\epsilon_{ij}^{\,\,\,\,k}\tau_k$ and $\{\tau_i,\tau_j\}=2\delta_{ij}$, it is straightforward to show:
\begin{equation}\label{eq:psp_simp}
    P_b\tau_zP_b = \vec{m}_b\cdot\vec{\tau} - \frac{1}{2}\cos\theta_b,
\end{equation}
where
\begin{equation}
    \vec{m}_b = (\sin2\theta_b\cos\phi_b,\sin2\theta_b\sin\phi_b,1+\cos2\theta_b).
\end{equation}
Using Eq.~\eqref{eq:psp_simp} and setting $\mu\pm\lambda>0$ and, hence, $\cos\theta_b>0$ for $b=U,L$ without loss of generality, it follows immediately that the eigenvalues of $\mathcal{\hat{S}}^z_{\,\,\mathcal{P}}$ are
\begin{equation}\label{eq:psp_eig}
    \lambda_{\mathcal{\hat{S}}^z_{\,\,\mathcal{P}}} = 0,\,-\cos\theta_U,\,\cos\theta_L.
\end{equation}
In Eq.~\eqref{eq:psp_eig}, $\lambda_{\mathcal{\hat{S}}^z_{\,\,\mathcal{P}}}=0$ are doubly degenerate and correspond to unoccupied states. Since $\cos\theta_b\neq0$  for $b=U,L$ everywhere in the $k$-space, the spectrum of $\mathcal{\hat{S}}^z_{\,\,\mathcal{P}}$ has a gap between sectors with positive and negative values of $\lambda_{\mathcal{\hat{S}}^z_{\,\,\mathcal{P}}}$. Therefore, from Eqs.~\eqref{eq:counter_ex_psp} and \eqref{eq:psp_simp}, we can define the projection operators onto the states on these sectors, as:
\begin{align}
    P_+ & = \begin{pmatrix}
    \frac{1}{2}(\mathbf{1}-\frac{\vec{m}_U}{\|\vec{m}_U\|}\cdot\vec{\tau}) & 0_{2\times2} \\ 0_{2\times2} & 0_{2\times2}
\end{pmatrix},
\\ P_- & = \begin{pmatrix}
    0_{2\times2} & 0_{2\times2} \\ 0_{2\times2} & \frac{1}{2}(\mathbf{1}-\frac{\vec{m}_L}{\|\vec{m}_L\|}\cdot\vec{\tau})
\end{pmatrix}.
\end{align}
$P_\pm$ have eigenvalues of $\hat{\Xi}$ equal $\pm1$, as $\Xi=\xi_z$ in the basis $(\hat{c}_{a\uparrow},\hat{c}_{b\downarrow},\hat{c}_{a\downarrow},\hat{c}_{b\uparrow})$. Since $P_\pm$ locate in different blocks, $P_\pm\partial^{\mu}P_\mp=0$ and, hence, $K_c=0$, even though $K\neq0$ and the spin-$U(1)$ symmetry is broken. 

\section{S4. Proof of $\text{tr}[G_{c,+}]=\text{tr}[G_{c,-}]$}\label{sec:S3}
\par We prove that $\text{tr}[G_{c,+}]=\text{tr}[G_{c,-}]$, where $\pm$ denotes the only two sectors in the projected spectrum with positive and negative eigenvalues. We have,
\begin{align}
\text{tr}[G_{c,-}] = \sum_{\mu=x,y}\text{Tr}\left[\partial^\mu P_- P_+ \partial^\mu P_-\right] = \sum_{\mu=x,y}\sum_{n\in-}\sum_{m\in+}|\braket{\tilde{u}_m^{(+)}|\partial^\mu\tilde{u}_n^{(-)}}|^2.
\end{align}
Also, since $\braket{\tilde{u}_m^{(+)}|\tilde{u}_n^{(-)}}=0$, we obtian
\begin{equation}
\braket{\tilde{u}_n^{(+)}|\partial^\mu\tilde{u}_n^{(-)}} = - \braket{\partial^\mu\tilde{u}_n^{(+)}|\tilde{u}_n^{(-)}} = - (\braket{\tilde{u}_n^{(-)}|\partial^\mu\tilde{u}_n^{(+)}})^*.
\end{equation}
Therefore,
\begin{equation}
\text{tr}[G_{c,-}] = \sum_{\mu=x,y}\sum_{n\in-}\sum_{m\in+}|\braket{\tilde{u}_m^{(+)}|\partial^\mu\tilde{u}_n^{(-)}}|^2 = \sum_{\mu=x,y}\sum_{n\in-}\sum_{m\in+}|\braket{\tilde{u}_n^{(-)}|\partial^\mu\tilde{u}_m^{(+)}}|^2 = \sum_{\mu=x,y}\text{Tr}\left[\partial^\mu P_+ P_- \partial^\mu P_+\right] = \text{tr}[G_{c,+}].
\end{equation}

\section{S5. Optical origin of the topological bound}
\par We show that the topological bound in Eq.~(2) of the main text originates from the positivity of optical absorption power for systems characterized by a narrow occupied-state bandwidth $t$ and a large band gap $E_{\mathrm{gap}}$. Following the discussion in Ref.~\cite{PhysRevX.14.011052}, the absorbed power of circularly polarized light is given by
\begin{equation}
    P_{\pm} = \left[\mathrm{Re}(\sigma_{xx}+\sigma_{yy}) \pm 2\,\mathrm{Im}(\sigma_{xy})\right] E^2,
\end{equation}
where $\pm$ denotes the helicity and $E$ is the electric-field amplitude of the circularly polarized light. Since $P_{\pm}\geq 0$ for both helicities, one obtains the inequality
\begin{equation}\label{eq:optical_ineq}
    \sum_{\mu=x,y}\int_{0}^{\infty} \! d\omega \,
    \frac{\mathrm{Re}\,\sigma_{\mu\mu}(\omega)}{\omega}
    \geq
    2\left|\int_{0}^{\infty} \! d\omega \,
    \frac{\mathrm{Im}\,\sigma_{xy}(\omega)}{\omega}\right|
    = \pi |\sigma_{xy}(0)|,
\end{equation}
where the last equality follows from the Kramers-Kronig relations. Equation~\eqref{eq:optical_ineq} can be further connected to quantum geometry, yielding \cite{PhysRevX.14.011052}
\begin{equation}\label{eq:optical_ineq_2}
    \sum_{\mu=x,y}\frac{\pi e^2}{\hbar}
    \int d[\mathbf{k}]\,
    \mathrm{Tr}\!\left[
        \partial^{\mu} P_{\mathrm{occ.}}\,
        Q\,
        \partial^{\mu} P_{\mathrm{occ.}}
    \right]
    \geq
    \frac{e^2}{2\hbar}\,|C|.
\end{equation}
Here, $P_{\mathrm{occ.}}$ denotes the projector onto the occupied states, $Q=1-P_{\mathrm{occ.}}$, and $C$ is the Chern number of the occupied manifold. In this sense, the optical measurement protocol discussed in the main text corresponds to a finite cutoff-frequency implementation of the right-hand sides of Eqs.~\eqref{eq:optical_ineq} and \eqref{eq:optical_ineq_2}.

\par We next consider a system with a narrow occupied-bandwidth $t$ and a large band gap $E_{\mathrm{gap}}$, satisfying $t \ll E_{\mathrm{gap}}$ to which we apply a Zeeman field with an associated energy scale $E_{h_z}$ obeying the hierarchy $t \ll E_{h_z} \ll E_{\mathrm{gap}}$. When the conduction bands are sufficiently dispersive such that $E_{h_z}$ is much smaller than their bandwidth, the unoccupied states can be regarded as effectively unchanged by the Zeeman field. Under these conditions, together with the situation depicted in Fig.~2 of the main text, applying Eq.~\eqref{eq:optical_ineq_2} yields:
\begin{equation}\label{eq:ineq_neg}
\sum_{\mu=x,y}\frac{\pi e^2}{\hbar}\int d[\mathbf{k}]\left(\text{Tr}[\partial^{\mu}P_{-}P_{+}\partial^{\mu}P_{-}]+ \text{Tr}[\partial^{\mu}P_{-}Q\partial^{\mu}P_{-}]\right) \geq \frac{e^2}{2\hbar}|C_{-}|.
\end{equation}
Here, $P_+$ ($P_-$) is the projection operator onto the states with positive (negative) $\mathcal{\hat{S}}^z_{\,\,\mathcal{P}}$ eigenvalues and $C_{\pm}$ is the corresponding Chern number. Similarly, applying the Zeeman field in the opposite direction yields: 
\begin{equation}\label{eq:ineq_pos}
\sum_{\mu=x,y}\frac{\pi e^2}{\hbar}\int d[\mathbf{k}]\left(\text{Tr}[\partial^{\mu}P_{+}P_{-}\partial^{\mu}P_{+}]+ \text{Tr}[\partial^{\mu}P_{+}Q\partial^{\mu}P_{+}]\right) \geq \frac{e^2}{2\hbar}|C_{+}|.
\end{equation}
Combining Eqs.~\eqref{eq:ineq_neg} and \eqref{eq:ineq_pos} yields:
\begin{equation}
    \sum_{\mu=x,y}2\pi\int d[\mathbf{k}]\left(\text{Tr}[\partial^{\mu}P_{+}P_{-}\partial^{\mu}P_{+}] + \text{Tr}[\partial^{\mu}P_{-}P_{+}\partial^{\mu}P_{-}] + \text{Tr}[\partial^{\mu}P_{+}Q\partial^{\mu}P_{+}] + \text{Tr}[\partial^{\mu}P_{-}Q\partial^{\mu}P_{-}]\right) \geq |C_{+}|+|C_{-}|,
\end{equation}
which is essentially the topological bound described in Eq.~(2) of the main text:
\begin{equation}
    K_c + K \geq |C_{+}|+|C_{-}|.
\end{equation}
While this section focuses on the SCIs characterized by the projected spin spectrum $\hat{\mathcal{S}}^{z}_{\mathcal{P}}$, the derivation can be straightforwardly extended to more general operators $\hat{O}$ whose projected spectra consist of two distinct sectors. Moreover, extensions to operators $\hat{O}$ with multiple sectors can be systematically constructed using the formalism developed in the next section. 

\section{S6. Comments on experimental implication of the topological bound for the case of $\mathcal{\hat{O}}_{\mathcal{P}}$ with multiple sectors}
\par We discuss how to experimentally verify the topological bound via the optical conductivity sum rule in systems with a chosen $\hat{O}$ that exhibits multiple sectors in its projected spectrum. As in the main text, we focus on systems with a narrow occupied-state bandwidth $t$ and a large band gap $E_{\text{gap}}\gg t$. We assume $\hat{O}$ enters the Hamiltonian through a tunable physical field $F_{\hat{O}}$ with energy scale $E_{F_{\hat{O}}}$, where $t\ll E_{F_{\hat{O}}}\ll E_{\text{gap}}$. $F_{\hat{O}}$ is applied such that the occupied states split across the Fermi level but remain well isolated from the unoccupied states. The corresponding low-energy Hamiltonian is $H_{\text{low-E}}\propto\mathcal{\hat{O}}_{\mathcal{P}}$. 
\begin{figure}[h]
  \centering
  \centering
    \includegraphics[width=0.6\linewidth]{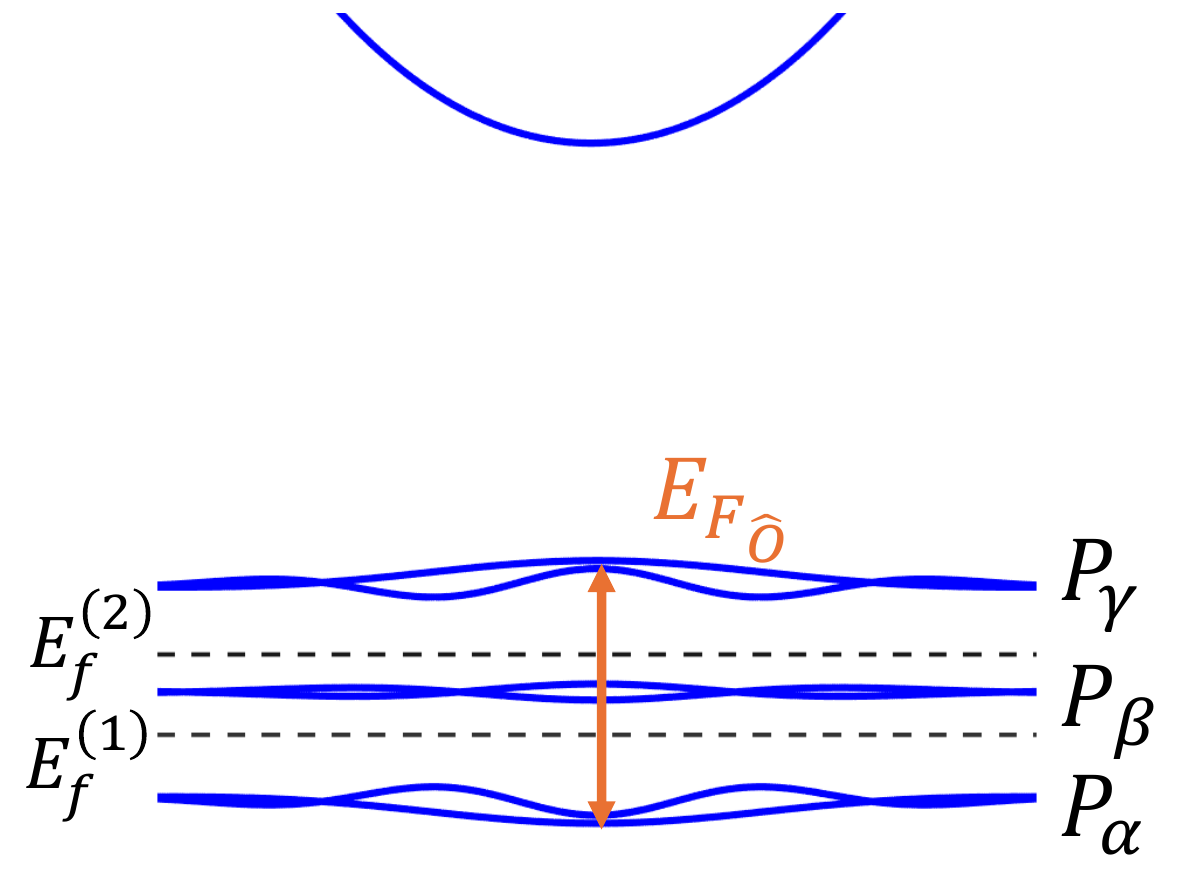}
  \caption{ Schematic band structure showing a narrow occupied-state bandwidth and a large band gap after applying a tunable physical field $F_{\hat{O}}$ with energy scale $E_{F_{\hat{O}}}$. Dashed lines mark the Fermi levels for two optical conductivity measurements.}
  \label{fig:measure_Kc_3}
\end{figure}

\par This setups enable the $K_c$ to be extracted via multiple optical conductivity measurements. As an illustrative example, we consider the case where $\mathcal{\hat{O}}_{\mathcal{P}}$ exhibits three sectors labeled by $\alpha$, $\beta$, and $\gamma$ (Fig.~\ref{fig:measure_Kc_3}). Measurement for the first sum rule proceeds by tuning the Fermi level $E_f^{(1)}$ to lie between sectors $\alpha$ and $\beta$ (Fig.~\ref{fig:measure_Kc_3}) and choosing a cutoff frequency $\Omega_1$ satisfying $E_{\alpha\beta}<\hbar\Omega_1<E_{\text{gap}}$ such that the optical conductivity sum rule yields:
\begin{equation}\label{eq:sum_rule_3_1}
W_{E_f^{(1)}}^{\mu\nu}\equiv\int_0^{\Omega_1} d\omega \frac{\sigma^{(\text{abs})}_{\mu\nu}(\omega)}{\omega} = \frac{\pi e^2}{\hbar} \int d[\mathbf{k}] \text{Tr}\left[\partial^\mu P_\alpha P_\beta \partial^\nu P_\alpha + \partial^\mu P_\alpha P_\gamma \partial^\nu P_\alpha \right].
\end{equation}
Here, $E_{\alpha\beta}=\text{max}(\{E_\beta\})-\text{min}(\{E_\alpha\})$ with $\{E_\alpha\}$ are the energies of states in the $\alpha$th sector, $E_f^{(i)}$ indicates the Fermi level for measuring the $i$th sum rule, and $P_\alpha$ is the projection operator onto states in the $\alpha$th sector. The measurement for the second sum rule proceeds by tuning the Fermi level $E_f^{(2)}$ to lie between sectors $\beta$ and $\gamma$ (Fig.~\ref{fig:measure_Kc_3}) and choosing a cutoff frequency $\Omega_2$ satisfying $E_{\beta\gamma}<\hbar\Omega_2<E_{\text{gap}}$ such that the optical conductivity sum rule yields:
\begin{equation}\label{eq:sum_rule_3_2}
W_{E_f^{(2)}}^{\mu\nu}\equiv\int_0^{\Omega_2} d\omega \frac{\sigma^{(\text{abs})}_{\mu\nu}(\omega)}{\omega} = \frac{\pi e^2}{\hbar} \int d[\mathbf{k}] \text{Tr}\left[\partial^\mu P_\beta P_\gamma \partial^\nu P_\beta \right].
\end{equation}
Note that $\text{Tr}\left[\partial^\mu P_\alpha P_\beta \partial^\mu P_\alpha \right] = \text{Tr}\left[\partial^\mu P_\beta P_\alpha \partial^\mu P_\beta \right]$ can be easily proved using methods discussed in Sec.S3. Therefore, combining Eqs.~\eqref{eq:sum_rule_3_1} and \eqref{eq:sum_rule_3_2}, we obtain:
\begin{align}
2\left(W_{E_f^{(1)}}^{\mu\mu} + W_{E_f^{(2)}}^{\mu\mu}\right) & = \frac{2\pi e^2}{\hbar} \int d[\mathbf{k}]  \text{Tr}\left[\partial^\mu P_\alpha P_\beta \partial^\mu P_\alpha + \partial^\mu P_\alpha P_\gamma \partial^\mu P_\alpha + \partial^\mu P_\beta P_\gamma \partial^\mu P_\beta \right], \notag
\\ & = \frac{\pi e^2}{\hbar} \int d[\mathbf{k}]  G^{\mu\mu}_{c,\alpha} + \text{Tr}\left[\partial^\mu P_\beta P_\alpha \partial^\mu P_\beta + \partial^\mu P_\gamma P_\alpha \partial^\mu P_\gamma + \partial^\mu P_\beta P_\gamma \partial^\mu P_\beta + \partial^\mu P_\gamma P_\beta \partial^\mu P_\gamma \right], \notag
\\ & = \frac{\pi e^2}{\hbar} \int d[\mathbf{k}]  G^{\mu\mu}_{c,\alpha} + G^{\mu\mu}_{c,\beta} + G^{\mu\mu}_{c,\gamma}, \notag
\\ & = \frac{\pi e^2}{\hbar} \int d[\mathbf{k}] G^{\mu\mu}_{c}. 
\end{align}
$K_c$ can thus be extracted via the optical conductivity sum rules with different cutoff frequencies by measuring $\text{tr}[\sigma^{(\text{abs})}(\omega)]$ using linearly polarized light at different Fermi levels:
\begin{equation}
K_c = \frac{4\hbar}{e^2}\text{tr}[W_{E_f^{(1)}} + W_{E_f^{(2)}}] = \frac{4\hbar}{e^2} \left( \int_0^{\Omega_1} d\omega\frac{\text{tr}[\sigma^{(\text{abs})}(\omega)]}{\omega} + \int_0^{\Omega_2} d\omega\frac{\text{tr}[\sigma^{(\text{abs})}(\omega)]}{\omega} \right).
\end{equation}
Since $K$ can also be extracted via the optical conductivity sum rules before applying the $F_{\hat{O}}$ field, the topological bound can therefore be experimentally verified. An example is a trilayer of coupled Chern insulators, where $\hat{O}$ corresponds to the layer pseudospin $\tau_z=\text{diag}(1_{n\times n},0_{n\times n},-1_{n\times n})$ with $n$ be the number of electrons in a layer and $F_{\hat{O}}$ is an out-of-plane electric field.

\par Although we only discussed $\mathcal{\hat{O}}_{\mathcal{P}}$ exhibiting three sectors, the procedure can be easily generalized. $K_c$ and, hence, the topological bound in systems with $\mathcal{\hat{O}}_{\mathcal{P}}$ exhibiting $m$ sectors can be experimentally verified by utilizing $m-1$ optical conductivity sum rules with different cutoff frequencies under the field $F_{\hat{O}}$, each obtained by measuring $\text{tr}[\sigma^{(\text{abs})}(\omega)]$ at different Fermi levels using linearly polarized light.

\end{document}